\documentclass[11pt]{amsart}

\usepackage[top=0.95in,bottom=1.1in,left=1.05in,right=1.05in]{geometry}

\usepackage{amsmath,amssymb,amsthm,bbm,bm,soul}
\usepackage{verbatim}
\usepackage{latexsym}
\usepackage{indentfirst}
\usepackage{mathrsfs}
\usepackage{placeins}
\usepackage{algorithm,algorithmic}
\usepackage[hidelinks]{hyperref}
\usepackage{fancybox}
\usepackage{fancyhdr}
\usepackage{graphicx}
\usepackage{color}
\usepackage{float}

\numberwithin{equation}{section}

\newcommand{\R}{\mathbb{R}}
\newcommand{\RR}{\mathbb{R}}
\newcommand{\PP}{\mathbb{P}}
\newcommand{\EE}{\mathbb{E}}
\newcommand{\E}{\mathbb{E}}

\newcommand{\XX}{\mathcal{X}}

\newcommand{\TT}{\mathcal{T}}

\newcommand{\FF}{\mathcal{F}}
\newcommand{\GG}{\mathcal{G}}
\newcommand{\HH}{\mathcal{H}}
\newcommand{\cK}{\mathcal{K}}
\newcommand\myf{f}

\newcommand{\ZZ}{\mathcal{D}}

\newcommand{\eps}{\epsilon}
\newcommand{\la}{\lambda}

\newcommand{\mk}[1]{\mathfrak{#1}}
\newcommand{\myM}{\mathcal{M}}

\newcommand{\red}[1]{{#1}}
\DeclareMathOperator\diag{diag}
\DeclareMathOperator\arginf{arg\ inf}
\DeclareMathOperator\argsup{arg\ sup}
\DeclareMathOperator\sign{sign}

\DeclareMathOperator*{\sspan}{span}

\newtheorem{theo}{Theorem}[section]

\newtheorem{remark}[theo]{Remark}

\begin{document}
\begin{abstract}
We investigate two new strategies for the numerical solution of optimal stopping problems within the Regression Monte Carlo (RMC) framework of Longstaff and Schwartz. First, we propose the use of stochastic kriging (Gaussian process) meta-models for fitting the continuation value. Kriging offers a flexible, nonparametric regression approach that quantifies approximation quality. Second, we connect the choice of stochastic grids used in RMC to the Design of Experiments paradigm. We examine space-filling and adaptive experimental designs; we also investigate the use of batching with replicated simulations at design sites to improve the signal-to-noise ratio. Numerical case studies for valuing Bermudan Puts and Max-Calls under a variety of asset dynamics illustrate that our methods offer significant reduction in simulation budgets over existing approaches.
\end{abstract}

\title{Kriging Metamodels and Experimental Design for Bermudan Option Pricing}
\author{Mike Ludkovski}
\address{Department of  Statistics and Applied Probability\\
  University of California, Santa Barbara \url{ludkovski@pstat.ucsb.edu}}
\thanks{Work Partially Supported by NSF CDSE-1521743}
\maketitle

\section{Introduction}

The problem of efficient valuation and optimal exercise of American/Bermudan options remains one of intense interest in computational finance. The underlying timing flexibility, which is mathematically mapped to an optimal stopping objective, is ubiquitous in financial contexts, showing up both directly in various derivatives and as a building block in more complicated contracts, such as multiple-exercise opportunities or sequential decision-making. Since timing optionality is often embedded on top of other features, ideally one would have access to a flexible pricing architecture to easily import optimal stopping sub-routines. While there are multiple alternatives that have been proposed and investigated, most existing methods are not fully scalable across the range of applications, keeping  the holy grail of such ``black-box'' optimal stopping algorithm out of reach\footnote{\red{By black-box we mean a ``smart'' framework that automatically adjusts low-level algorithm parameters based on the problem setting. It might still require some high-level user input, but avoids extensive fine-tuning or, at the other extreme, a hard-coded, non-adaptive method.}}. This is either due to reliance on approaches that work only for a limited subset of models, (e.g.~one-dimensional integral equations, various semi-analytic formulas, etc.) or due to severe curses of dimensionality that cause rapid deterioration in performance as model complexity increases.

Within this landscape, the regression Monte Carlo framework or RMC (often called Least Squares Monte Carlo, though see our discussion on this terminology in Section \ref{sec:critique}) has emerged as perhaps the most popular \emph{generic} method to tackle optimal stopping. The intrinsic scalability of Monte Carlo implies that if the problem is more complex one simply runs more simulations, with the underlying implementation  essentially independent of dimensionality, model dynamics, etc. However, the comparatively slow convergence rate of RMC places emphasis on obtaining more accurate results with fewer simulated paths, spurring an active area of research, see e.g.~\cite{AgarwalSircar15,BelomestnyNagapetyan15,GobetTurkedjiev15,Hepperger13,JainOosterlee13,LetourneauStentoft14,TompaidisYang13}.

In this article, we propose a novel marriage of modern statistical frameworks and the optimal stopping problem, making two methodological contributions. First, we examine the use of kriging meta-models as part of a simulation-based routine to approximate the optimal stopping policies. \red{Kriging offers a flexible non-parametric method (with several additional convenient features discussed below) for approximating the continuation value, and as we demonstrate is competitive with existing basis-expansion setups.} To our knowledge this is the first article to use kriging models in optimal stopping. Second, we investigate the experimental design aspect of RMC. We propose several alternatives on how to generate and customize the respective stochastic grids, drawing attention to the accompanying performance gains. Among others, we investigate adaptive generation of the simulation designs, extending the ideas in Gramacy and Ludkovski \cite{GL13} who originally proposed the use of sequential design for optimal stopping.
\red{Rather than purely targeting speed savings, our strategies explore opportunities to reduce the simulation budget of RMC through ``smart'' regression and design approaches. In particular, the combination of kriging and improved experimental design brings up to an order-of-magnitude savings in simulation costs, which can be roughly equated to memory requirements.  While there is considerable overhead added,  these gains indicate the
potential of these techniques to optimize the speed-memory trade-off. They also show the benefits of approaching RMC as an iterated meta-modeling/response-surface modeling problem. }

Our framework is an outgrowth of the recent work by the author in \cite{GL13}. In relation to the latter paper, the present article makes several modifications that are attractive for the derivative valuation context. First, while \cite{GL13} suggested the use of piecewise-defined Dynamic Trees regression, the kriging meta-models are intrinsically continuous. As such, they are better suited for the typical financial application where the value function is smooth. Second, to reduce computational overhead, we introduce a new strategy of \emph{batching} by generating multiple scenarios for a given initial condition. Third, in contrast to \cite{GL13} that focused on sequential designs, we also provide a detailed examination and comparison of various \emph{static} experimental designs. Our experiments indicate that this already achieves significant simulation savings with a much smaller overhead, \red{and is one of the ``sweet spots'' in terms of overall performance.}

The rest of the paper is organized as follows. Section \ref{sec:model} sets the notation we use for a generic optimal stopping problem, and the RMC paradigm for its numerical approximation. Along the way we recast RMC as a sequence of meta-modeling tasks.  Section \ref{sec:kriging} introduces and discusses stochastic kriging meta-models that we propose to use for empirical fitting of the continuation value. Section \ref{sec:designs} then switches to the second main thrust of this article, namely the issue of experimental designs for RMC. We investigate space-filling designs, as well as the idea of batching or replication.  Section \ref{sec:sequential} marries the kriging methodology with the framework of sequential design to obtain an efficient approach for generating designs adapted to the loss criterion of RMC. In Section \ref{sec:examples} we present a variety of case studies, including a classical bivariate GBM model, several stochastic volatility setups, and multivariate GBM models with multiple stopping regions. Finally, Section \ref{sec:conclude} summarizes our proposals and findings.

\section{Model}\label{sec:model}
We consider a discrete-time optimal stopping problem on a finite horizon. Let  $(X_t)$, $t=0,1,\ldots,T$ be a Markov state process on a stochastic basis $(\Omega, \mathcal{F}_\infty, \PP)$ taking values in  some (usually uncountable) subset $\XX \subseteq \mathbb{R}^d$.  The financial contract in question has a maturity date of $T < \infty$ and can be exercised at any earlier date with payoff $h(t,x) \in \RR$. Note that in the financial applications this corresponds to a Bermudan contract with a unit of time corresponding to the underlying discretization $\Delta t= 1$ of early exercise opportunities. The dependence of $h(\cdot,x)$ on $t$ is to incorporate interest rate discounting.

Let $\FF_t = \sigma( X_{1:t})$, be the information filtration generated by $(X_t)$ and $\mathcal{S} $ the collection of all $(\FF_t)$-stopping times bounded by  $T$.
The Bermudan contract valuation problem consists of maximizing
the expected reward $h(\tau,X_\tau)$ over all $\tau \in \mathcal{S}$. More precisely, define for any $0 \le t \le T$,
\begin{align}\label{eq:Vt}
V(t,x) := \sup_{\tau \ge t, \tau \in \mathcal{S}} \EE_{t,x} \left[ h( \tau, X_\tau) \right],
\end{align}
where $\EE_{t,x}[\cdot] \equiv \EE[ \cdot | X_t = x]$ denotes $\PP$-expectation given initial condition $x$. We assume that $h(t,\cdot)$ is such that $V(t,x) < \infty$ for any $x$, for instance bounded.

Using the tower property of conditional expectations and one-step analysis,
\begin{align}\notag
V(t,x) &= \sup_{\tau \ge t, \tau \in \mathcal{S}} \EE[ h(\tau, X_\tau) | X_t=x] \\
 & = \max \left( h(t,x), C(t+1, X_{t+1}) \right) = h(t,x) + \max( T(t,x), 0), \label{eq:recurse-V}
\end{align}
where we defined the continuation value $C(t,x)$ and timing-value $T(t,x)$ via
\begin{align} \label{def:C}
C(t,x) &:= \EE_{t,x} \left[ V(t+1,X_{t+1}) \right];
 \\ \label{def:T} T(t,x) &:= C(t,x) - h(t,x).
\end{align}
Since $V(t,x) \ge h(t,x)$ for all $t$ and $x$, the smallest optimal stopping time $\tau^*(t,x)$ for \eqref{eq:Vt} satisfies
\begin{align}\label{eq:sign-of-T}
\{ \tau^*(t,x) = t \} = \left\{ h(t,x) \ge C(t,x) \right\} = \{ T(t,x) \le 0 \}.
\end{align}
Thus, it is optimal to stop immediately if and only if the conditional expectation of tomorrow's reward-to-go is less than
the immediate reward. Rephrasing, figuring out the decision to stop at $t$
is equivalent to finding the zero level-set (or contour) of the timing value $T(t,x)$ or {classifying} the state
space $\mathcal{X} \ni X_t$ into the \emph{stopping region} $\mk{S}_t := \{ x : T(t,x) \le 0 \}$ and its complement the continuation
region.
 By induction, an optimal stopping time is
\begin{align}\label{eq:first-hit}
\tau^*(t,x) = \inf \{s \ge t : X_s \in \mk{S}_s \} \wedge T.
\end{align}
From the financial perspective, obtaining the exercise strategy as defined by $\tau^*$ is often even more important than finding the present value of the contract.

The representation \eqref{eq:first-hit} suggests a recursive procedure to estimate the optimal policy as defined by $\tau^*$. To wit, given a collection $\widehat{\mk{S}}_{t+1:T}$ of subsets of $\XX$ which are stand-ins for the stopping regions, define the corresponding exercise strategy pathwise for a scenario $\omega$:
\begin{align}\label{eq:hat-tau}
  \hat{\tau}(t,x) (\omega) := \inf \{ s > t: X_s(\omega) \in \widehat{\mk{S}}_s \} \wedge T.
\end{align}
Equipped with $\hat{\tau}$, we obtain the pathwise payoff
\begin{align}\label{eq:H}
H_t( X^x_\cdot; \widehat{\mk{S}}_{t+1:T})) ( \omega) := h(  \hat{\tau}(t,x) (\omega), X_{\hat{\tau}(t,x) (\omega)}(\omega) ).
\end{align}
Taking expectations, $\E_{t,x}[ H_t(X^x_\cdot; \widehat{\mk{S}}_{t+1:T})]$ is the expected payoff starting at $x$, \emph{not} exercising immediately at step $t$ and then following the exercise strategy specified by $\widehat{\mk{S}}_{t+1:T}$; our notation highlights the latter (path-) dependence of $H_t$ on future stopping regions $\widehat{\mk{S}}_t$'s.
 If $\widehat{\mk{S}}_s = \mk{S}_s$ for all $s > t$, we can use \eqref{eq:H} to recover $V(t,x)$. Indeed, the expected value of $H_t$ in this case is precisely the continuation value $\EE_{t,x}[ H_t( X^x_\cdot; \mk{S}_{t+1:T} ) ] = C(t,x)$. Consequently, finding $C(t,x)$ (and hence the value function via \eqref{eq:recurse-V}) is reduced to
 computing conditional expectations of the form $x \mapsto \EE_{t,x}[ H_t(X^x_\cdot)]$. Because analytic evaluation of this map is typically not tractable, this is the starting point for Monte Carlo approximations which only require the ability to simulate $(X_t)$-trajectories. In particular, simulating $N$ independent scenarios $x^{n}_{t:T}$, $n=1,\ldots,N$ emanating from $x^{n}_t = x$ yields the estimator ${\hat{V}}(t,x)$ via
\begin{align}\label{eq:mc-average}
  {\widehat{C}}(t,x) := \frac{1}{N} \sum_{n=1}^N H_t( x^{n}_\cdot), \qquad {\hat{V}}(t,x) := \max( h(t,x),  {\widehat{C}}(t,x) ).
\end{align}
 A key observation is that the approximation quality of \eqref{eq:mc-average} is driven by the accuracy of the stopping sets $\widehat{\mk{S}}_{t+1:T}$ that determine the exercise strategy and control the resulting errors (formally we should thus use a double-hat notation to contrast the estimate in \eqref{eq:mc-average} with the true expectation $\hat{C}^*(t,x) =\E_{t,x}[ H_t(X^x_\cdot; \widehat{\mk{S}}_{t+1:T})]$).  Thus, for the sub-problem of  approximating the conditional expectation of $H_t(X_\cdot)$, there is a thresholding property, so that regression errors are tolerated as long as they do not affect the \emph{ordering} of $\EE_{t,x}[ H_t( X_\cdot) ]$ and $h(t,x)$. This turns out to be a major advantage compared to value function approximation techniques as in \cite{TsitsiklisVanRoy}, which use \eqref{eq:recurse-V} directly and do not explicitly make use of pathwise payoffs. \red{See also \cite{Stentoft14} for a recent comparison of value-function and stopping-time approximations. }

\subsection{Meta-Modeling}
In \eqref{eq:mc-average} the estimate of $\EE_{t,x}[ H_t( X^x_\cdot) ]$ was obtained pointwise for a given location $(t,x)$ through a plain Monte Carlo approach. Alternatively, given paths $x^{n}_{t:T}$, $n=1,\ldots,N$ emanating from \emph{different} initial conditions $x^{n}_t$, one may borrow information \emph{cross-sectionally} by employing a statistical regression framework. A regression model specifies the link
\begin{align}\label{eq:regress}
 H_t(X^x_\cdot) =C(t,x) + \eps(t,x)
\end{align}
where $\eps(t,x)$ is the mean-zero noise term with variance $\sigma^2(t,x)$. This noise arises from the random component of the pathwise payoff due to the internal fluctuations in $X_{t:T}$ and hence $\hat\tau(t,x)$. Letting $y^{1:N}_t \equiv H_t( x^{1:N}_\cdot)$ be a vector of obtained pathwise payoffs, one regresses $y^{1:N}_t$ against $x^{1:N}_t$ to fit an approximation $\hat{C}(t,\cdot)$. Evaluating $\hat{C}(t,\cdot)$ at any $x \in \XX$ then yields the predicted continuation value from the exercise strategy $\widehat{\mk{S}}_{t+1:T}$ conditional on $X_t= x$.

The advantage of regression is that it replaces pointwise estimation (which requires re-running additional scenarios for each location $x$) with a single step that fuses all information contained in $(x_t,y_t)^{1:N}$ to fit $\hat{C}(t,\cdot)$. Armed with the estimated $\hat{C}(t,\cdot)$, we take the natural estimator of the stopping region at $t$,
\begin{align}\label{eq:hatS-t}
  \widehat{\mk{S}}_t := \{ x : \hat{C}(t,x) \le h(t,x) \},
\end{align}
which yields an inductive procedure to approximate the full exercise strategy via \eqref{eq:hat-tau}. Note that \eqref{eq:hatS-t} is a double approximation of $\mk{S}_t$ -- due to the discrepancy between $\hat{C}(t,\cdot)$ and the true expectation $\E_{t,x}[H_t(X^x_\cdot; \widehat{\mk{S}}_{t+1:T})]$ from \eqref{eq:H}, as well as due to partially propagating the previous errors in $\widehat{\mk{S}}_{t+1:T}$ relative to the true $\mk{S}_{t+1:T}$.

The problem of obtaining $\hat{C}(t,\cdot)$ based on \eqref{eq:regress} is known as \emph{meta-modeling} or emulation in the machine learning and simulation optimization literatures \cite{Kleijnen,NelsonStaum10,SantnerBook}. It consists of two fundamental steps: first a design (i.e.~a grid) $\ZZ := x^{1:N}$ is constructed and the corresponding pathwise payoffs $y^{1:N}$ are realized via simulation. Then, a regression model \eqref{eq:regress} is trained to link the $y$'s to $x$'s via an estimated  $\hat{C}(t,\cdot)$.
In the context of \eqref{def:C}, emulation can be traced to the seminal works of Tsitsiklis and van Roy \cite{TsitsiklisVanRoy} and Longstaff and Schwartz \cite{LS} (although the idea of applying regression originated earlier, with at least Carri\`ere \cite{Carriere96}). The above references pioneered classical least-squares regression (known by the acronyms Least Squares Method, Least Squares Monte Carlo and more generally as Regression Monte Carlo and Regression Based Schemes) for \eqref{eq:regress}, i.e.~the use of projection onto a finite family of basis functions $\{ B_r(x) \}_{r=1}^R$,
$$
  \hat{C}(t,x) = \sum_{r=1}^R \beta_r B_r(x).
$$
\red{The projection was carried out by minimizing the $L^2$ distance between $\hat{C}(t,\cdot)$ and the manifold $\HH_R := \sspan( B_r )$} spanned by the basis functions (akin to the definition of conditional expectation):
\begin{align}\label{eq:ols}
\hat{C}(t,\cdot) = \arginf_{f \in \mathcal{H}_R} L_{2}( f),
\end{align}
where the loss function $L_{2}$ is a weighted $L^2$ norm based on the distribution of $X_t$,
\begin{align}\label{eq:weighted-norm}
L_{2}(f) = \| H_t-f \|^2_{L^2(X_t)} = \EE_{0,X_0} \left[ (H_t(X_\cdot)-f(X_t))^2 \right].
\end{align}

Motivated by the weights in \eqref{eq:weighted-norm}, the accompanying design $x^{1:N}$ is commonly generated by i.i.d.~sampling from $p(t,\cdot|0,X_0)$. As suggested in the original articles \cite{LS,TsitsiklisVanRoy} this can be efficiently implemented by a \emph{single} global simulation of $X$-paths $x^{1:N}_{0:T}$, although at the cost of introducing further $t$-dependencies among $\hat{C}(t,\cdot)$'s.

Returning to the more abstract view, least-squares regression \eqref{eq:ols} is an instance of a meta-modeling technique. Moreover, it obscures the twin issue of generating  $\ZZ = x^{1:N}$. Indeed,
in contrast to standard statistical setups, in meta-modeling there is no a priori \emph{data} per se; instead the solver is responsible both for generating the data and for training the model. These two tasks are intrinsically intertwined. The subject of Design of Experiments (DoE)  has developed around this issue, but has been largely absent in RMC approaches. In the next Section we re-examine existing RMC methods and propose novel RMC algorithms that build on the DoE/meta-modeling paradigm by targeting \emph{efficient} designs $\ZZ$.

\subsection{Closer Look at the RMC Regression Problem}\label{sec:critique}

\begin{figure}[ht]
  \centering
  \includegraphics[width=0.44\textwidth,height=2.2in,trim=0.15in 0.15in 0.15in 0.15in]{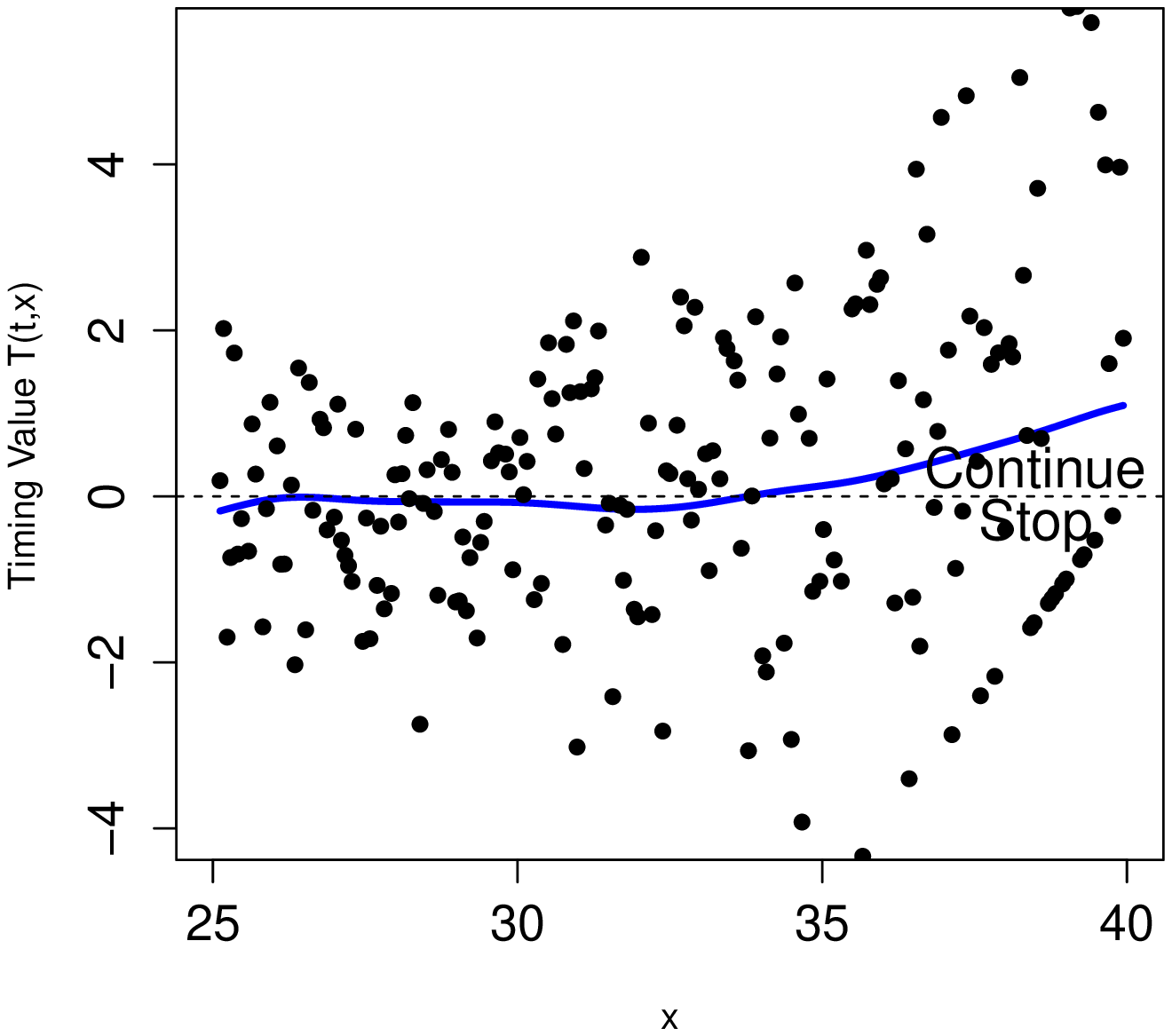}
 \includegraphics[width=0.44\textwidth,height=2.2in,trim=0.15in 0.15in 0.15in 0.15in]{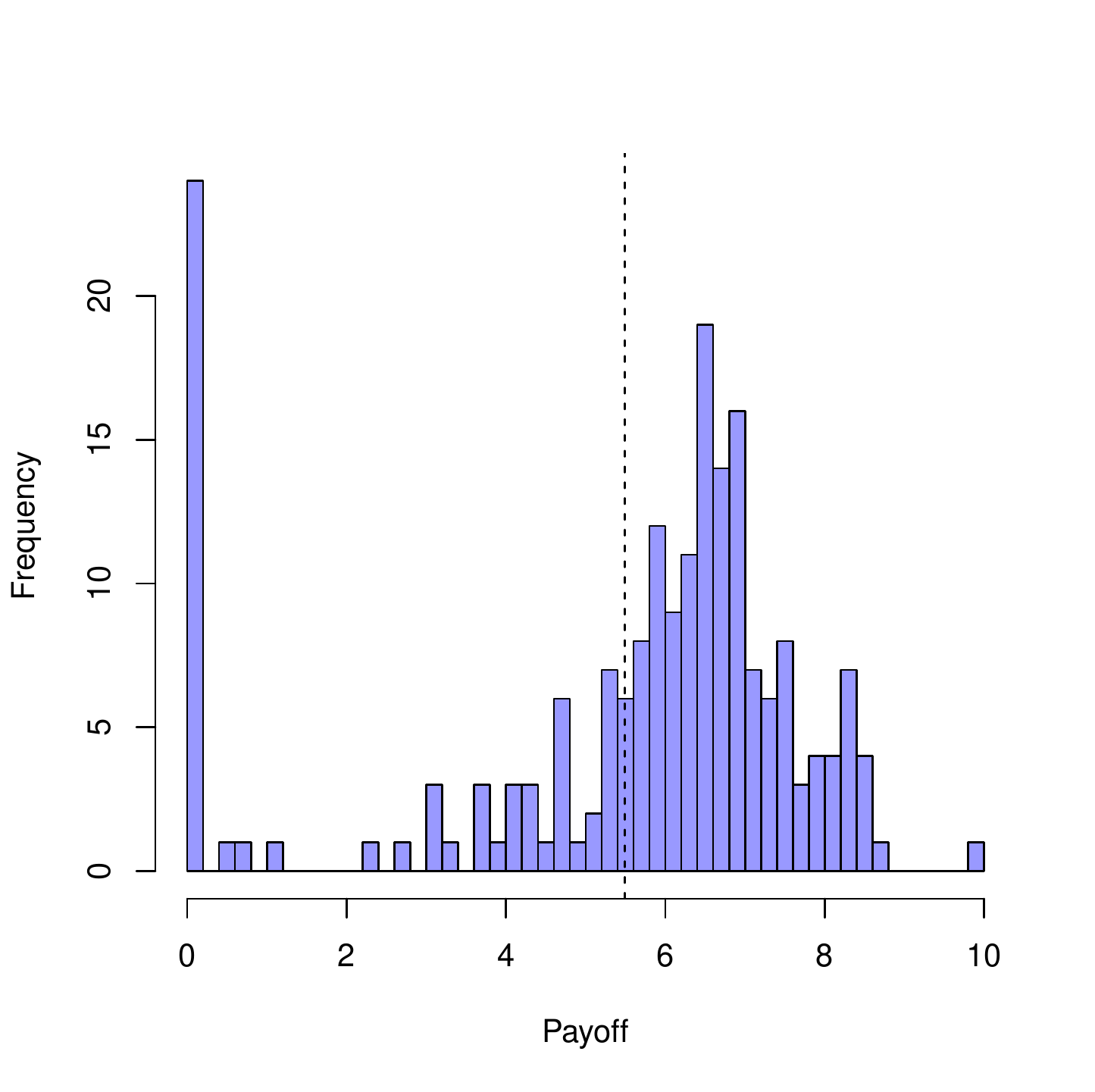}
  \caption{Pathwise payoffs in a 1-D Bermudan Put problem at $t=0.6$. \red{The underlying problem setting is described in Section \ref{sec:1d}}. Left: scatterplot of $(x, H_t(X^x_\cdot)-h(t,x))$ over 200 distinct $x \in \R_+$ (generated as a Sobol sequence). Right: Histogram of $N=200$ pathwise future payoffs $y^{1:N} \sim H_t(X^x_\cdot)$ starting at $x=35$. The vertical dashed line indicates the empirical mean $\EE[ H_t(X^x_\cdot) |X_t=35] \simeq Ave( y^{1:N}) =5.49$. In 24 out of 200 scenarios the payoff was zero, creating a point mass in the distribution of $H_t(X^x_\cdot)$. Other summary statistics were $StDev(y^{1:N}) = 2.45$, $Skew(y^{1:N}) = -1.28$ and $Max(y^{1:N}) = 9.87$.  \label{fig:payoffHist} }
\end{figure}

Figure \ref{fig:payoffHist} shows the distribution of $H_t(X^x_\cdot)$ in a 1-D Geometric Brownian motion  Bermudan Put option problem (see Section \ref{sec:1d}), the archetype of a financial application. The left plot shows a scatterplot of $(x,H_t(X^x_\cdot)-h(t,x))$ as $x \in \R_+$ varies, and the right panel gives a histogram of $H_t(X^x_\cdot)$ for a fixed initial value $X_t = x$. Two features become apparent from these plots. First, the noise variance $\sigma^2(t,x)$ is extremely large, swamping the actual shape of $x \mapsto T(t,x)$. Second, the noise distribution $\eps(t,x)$ is highly non-standard. It involves a potent brew of (i) heavy-tails; (ii) significant skewness; (iii) multi-modality. In fact, $\eps(t,x)$ does not even have a smooth density as the nonnegativity of the payoff $h(t,\cdot) \ge 0$, implies that $H_t(X_\cdot)$ has a point mass at zero.

Moreover, as can be observed in the left panel of Figure \ref{fig:payoffHist}, the distribution of $\eps(t,\cdot)$ is heteroscedastic, with a state-dependent conditional skew and state-dependent point mass at zero.
\red{These phenomena violate the assumptions of ordinary least squares regression, making the sampling distribution of fitted $\hat{\beta}_r$ ill-behaved, i.e.~far from Gaussian. More robust versions of \eqref{eq:ols}, including regularized (such as Lasso) or localized (such as Loess or piecewise linear models) least squares frameworks have therefore been proposed, see \cite{Kohler10review,Kohler12lasso}.} Further structured regression proposals can be found in \cite{KanReesor12,LetourneauStentoft14}. Alternatively, a range of variance reduction methods, such as control variates, have been suggested to ameliorate the estimation of $\hat{\beta}$ in \eqref{eq:ols}, see for example \cite{AgarwalSircar15,BelomestnyNagapetyan15,Hepperger13,JainOosterlee13,JunejaKalra09}.

\red{The parametric constraints placed by the LSMC approach \eqref{eq:ols} on the shape of the continuation value  $\hat{C} \in \HH_R$, make its performance highly sensitive to the distance between the  manifold $\mathcal{H}_R$ and the true $C(t,\cdot)$ \cite{Stentoft}.  
Moreover, when using \eqref{eq:ols} a delicate balance must be observed between the number of scenarios $N$ and the number of basis functions $R$. These concerns highlight the inflexible nature of parametric regression and become extremely challenging in higher dimensional settings with unknown geometry of $C(t,\cdot)$. One  work-around is to employ hierarchical representations, for example by partitioning $\XX$ into disjoint bins and employing independent linear hyperplane fits in each bin, see e.g.~Bouchard and Warin \cite{BouchardWarin10}. Nevertheless, there remain ongoing challenges in adaptive construction of $\mathcal{H}_R$.}

Another limitation of the standard LSMC approach is its objective function. As discussed, least squares regression can be interpreted as minimizing a global weighted mean-squared-error loss function. However, as shown in \eqref{eq:sign-of-T}, the principal aim in optimal stopping is not to learn $C(t,\cdot)$ globally in the sense of its mean-squared-error, but to rank it vis-a-vis $h(t,\cdot)$.
Indeed, recalling the definition of the timing function, the correct loss function $L_{RMC}$ (cf.~\cite{GL13})  is
\begin{align}\label{eq:loss}
L_{RMC}(\hat{C}) = \E_{0,X_0} \left[ |T(t,X_t)| 1_{\{\sign T(t,X_t) \neq \sign \hat{T}(t,X_t)\}} \right], \qquad \hat{T}(t,x) = \hat{C}(t,x) - h(t,x).
\end{align}
Consequently, the loss criterion is localized around the contour $\partial \mk{S}_t = \{  \hat{C}(t,x) =h(t,x) \} = \{ \hat{T}(t,x) = 0\}$. Indeed, a good intuition is that optimal stopping is effectively a classification problem: in some regions of $\mathcal{X}$, the sign of $T(t,x)$ is easy to detect and the stopping decision is ``obvious'', while in other regions $T(t,x)$ is close to zero and resolving the optimal decision requires more statistical discrimination. For example, in Bermudan options it is a priori  clear that out-of-the-money (OTM) $C(t,x)>h(t,x)$, and so it is optimal to continue at such $(t,x)$. This observation was already stated in \cite{LS} who suggested to only regress in-the-money (ITM) trajectories. However, it also belies the apparent inefficiency in the corresponding design--- the widely used LSMC implementation first ``blindly'' generates a design that covers the full domain of $X_t$, only to discard all OTM scenarios. This dramatically shrinks the effective design size, up to 80\% in the case of an OTM option. The mismatch between \eqref{eq:loss} and \eqref{eq:weighted-norm} makes the terminology of \emph{``least-squares''} Monte Carlo that exclusively emphasizes the $L^2$-criterion unfortunate.

\red{While one could directly use \eqref{eq:loss} during the cross-sectional regression (for example to estimate the coefficients $\hat{\beta}$ in \eqref{eq:ols}, which however would require implementing an entirely new optimization problem), a much more effective approach is to take this  criterion into account during the experimental design. Intuitively, the local accuracy of any approximation of $C(t,x)$  is limited by the amount of observations in the neighborhood, i.e.~the density of $\ZZ$ around $x$.
%
The form of \eqref{eq:loss} therefore suggests that the ideal approach would be to place $\ZZ$ along the stopping boundary $\partial \mk{S}_t$ in analogue to importance sampling. Of course the obvious challenge is that
knowing $\mk{S}_t$ is equivalent to solving the stopping problem. Nevertheless, this perspective suggests multiple improvements to the baseline approach which is entirely oblivious to \eqref{eq:loss} when picking $x^{1:N}$.}

\subsection{Outline of Proposed Approach}
\red{Motivated by the above discussion, we seek efficient meta-models for learning $C(t,\cdot)$ that can exploit the local nature of \eqref{eq:loss}. This leads us to consider non-parametric, kernel-based frameworks, specifically Gaussian process regression, also known as kriging \cite{NelsonStaum10,Kleijnen,WilliamsRasmussenBook}. We then
explore a range of Design of Experiments approaches for constructing $\ZZ$. Proposed design families include (i) uniform gridding; (ii) random and deterministic space-filling designs; (iii) adaptive sequential designs based on expected improvement criteria. While only the last choice is truly tailored to \eqref{eq:loss}, we explain that reasonably well-chosen versions of (i) and (ii) can already be a significant improvement on the baseline.}

To {formalize} construction of $\ZZ$, we rephrase \eqref{eq:regress} as a response surface (aka meta-) modeling (RSM) problem. RSM is concerned with the task of estimating an unknown function $x \mapsto \myf(x)$ that is noisily observed through a stochastic sampler,
\begin{align}\label{eq:resp-surface}
  Y(x) \sim \myf(x) + \eps(x), \qquad \EE[ \eps^2(x) ] = \sigma^2(x),
\end{align}
where we remind the reader to substitute in their mind $Y(x) \equiv H_t(X^x_\cdot)$ and $f \equiv C(t,\cdot)$;  $t$ is treated as a parameter.  Two basic requirements in RSM are a flexible nonparametric approximation architecture $\HH$ that is used for searching for the outputted fit $\hat{f}$, and global consistency, i.e.~convergence to the ground truth as number of simulations $N$ increases without bound. The experimental design problem then aims to maximize the information obtained from $N$ samples $(x,y)^{1:N}$ towards the end of minimizing the specified loss function $L(\hat{f})$.

\red{ The pathwise realizations of $H_t(X^x_\cdot)$ involve simulation noise which must be \emph{smoothed out} and as we saw is often highly non-Gaussian.} To alleviate this issue,  we
mimic the plain Monte Carlo strategy in \eqref{eq:mc-average} that uses the classical Law of Large Numbers to obtain a local estimate of $C(t,x)$, by constructing batched or replicated designs. Batching generates multiple independent samples $y^{(i)} \sim Y(x)$, $i=1,\ldots,M$ at the same $x$, which are used to compute the empirical average $\bar{y}= \frac{1}{M} \sum_i y^{(i)}$. Clearly, $\bar{Y}$ follows the same statistical model \eqref{eq:resp-surface} but with a signal-to-noise ratio improved by a factor of $M$, $\E[ \bar{\eps}^2(x)] = \sigma^2(x)/M$. This also reduces the post-averaged design size $| \ZZ|$, which speeds-up and improves the training of the meta-model.

\red{To recap, kriging gives a flexible non-parametric representation for $C(t,\cdot)$. The fitting method automatically detects the spatial structure of the continuation value obviating the need to specify the approximation function-class. While the user must pick the kernel family, extensive experiments suggest that these have a second-order effect.  Also, the probabilistic structure of kriging offers convenient goodness-of-fit diagnostics after $\hat{C}$ is obtained, yielding a tractable quantification of posterior uncertainty. This can be exploited for DoE, see Section \ref{sec:sequential}.}

\section{Kriging Metamodels}\label{sec:kriging}
Kriging models have emerged as perhaps the most popular framework  for DoE over continuous input spaces $\XX$. \red{In kriging the cross-sectional interpolation is driven by the spatial smoothness of $C(t,\cdot)$ and essentially consists of aggregating the observed values of $Y$ in the neighborhood of $x$.}  Book-length treatment of kriging can be found in \cite{WilliamsRasmussenBook}, see also \cite[Ch.~5]{Kleijnen} and \cite{NelsonStaum10}. To be precise, in this article we deal with stochastic kriging under heterogeneous sampling noise.

Stochastic
kriging meta-models follow a Bayesian paradigm, treating $\myf$ in \eqref{eq:resp-surface} as a random object belonging to a function space $\HH$.
Thus, for each $x$, $\myf(x)$ is a random variable whose posterior distribution is obtained based on the collected information from samples $(x, y)^{1:N}$ and its prior distribution $\GG_0$. Let $\GG$ be the information generated by the design $\ZZ$, i.e.~$\GG = \sigma( Y(x) : x \in x^{1:N} )$. We then define the posterior distribution $\myM_x(\cdot)$ of $\myf(x)$; the global map $x \mapsto \myM_x$ is called the surrogate surface (the terminology is a misnomer, since $\myM_x$ is in fact measure-valued).
Its first two moments are the surrogate mean and variance respectively,
\begin{align}
  m(x)  & := \EE[ \myf(x) | \GG] = \int_\R y \myM_x(dy),\\
v(x)^2 & :=\EE[ (\myf(x)-m(x))^2 | \GG] = \int_\R (y-m(x))^2 \myM_x(dy).
\end{align}
Note that in this function-view framework, a point estimate $\hat{f}$ of the unknown $\myf$ is replaced with a full posterior distribution $\EE[ f |\GG]$ over the space $\HH$; tractability is achieved by imposing and maintaining Gaussian structure. The surrogate mean surface $x\mapsto m(x)$ is then the natural proxy (being the maximum a posteriori probability estimator) for $\hat{f}$, and the surrogate variance $v^2(\cdot)$ is the proxy for the standard error of $m(\cdot)$.

To model the relationship between $f(x)$ and $f(x')$ at different locations $x,x' \in \XX$
kriging uses the Reproducing Kernel Hilbert Space (RKHS) approach, treating the full $f(\cdot)$ as a realization of a mean-zero Gaussian process. If necessary, $f$ is first ``de-trended'' to justify the mean-zero assumption. The Gaussian process generating $f$ is based on a  covariance kernel $\cK : \XX^2 \to \mathbb{R}$, with $\cK(x,x') = \EE[ \myf(x)\myf(x') | \GG_0]$. The resulting function class  $\HH_\cK := \sspan( \cK(\cdot, x'), x' \in \XX)$ forms a Hilbert space. The RKHS structure implies that both the prior and posterior distributions of $\myf(\cdot)$ given $\GG$ are multivariate Gaussian.

By specifying the correlation behavior, the kernel $\cK$ encodes the smoothness of the response surfaces drawn from the GP $\HH_{\cK}$ which is measured in terms of the RKHS norm $\| \cdot \|_{\HH_{\cK}}$. Two main examples we use are the squared exponential kernel
\begin{align}\label{eq:sqexp} \cK(x,x'; s, \vec{\theta}) = s^2 \exp \left( -  \|x -x'\|^2_\theta/2 \right),\end{align}
and the Matern-5/2 kernel
 \begin{align}\label{eq:matern}
 \cK(x,x'; s, \vec{\theta}) = s^2 \bigl( 1+ \sqrt{5}\| x-x'\|_\theta + \frac{5}{3} \| x-x' \|^2_\theta \bigr) \cdot e^{ - \sqrt{5} \| x - x'\|_\theta},
 \end{align}
which both use the weighted Euclidean norm $\|x\|^2_\theta = { x \cdot (\diag \vec\theta) \cdot x^T} = \sum_{j=1}^d \theta_j x^2_j $. The length-scale vector $\vec\theta$ controls the smoothness of members of $\HH_\cK$, the smaller the rougher.
The variance parameter $s^2$ determines the amplitude of fluctuations in the response. In the limit $\theta \to +\infty$, elements of $\HH_\cK$ concentrate on the \emph{linear} functions, effectively embedding linear regression; if $s=0$, elements of $\HH_\cK$ are constants. For both of the above cases, members of the function space $\HH_\cK$ can uniformly approximate any continuous function on any compact subset of $\XX$.

The use of different length-scales $\theta_j$ for different coordinates of $x$ allows anisotropic kernels that reflect varying smoothness of $f$ in terms of its different input coordinates. For example, in Bermudan option valuation, continuation values would typically be more sensitive to the asset price $X^1$ than to a stochastic volatility factor $X^2$, which would be reflected in $\theta_1 < \theta_2$.

Let $\vec{y} = (y(x^{1}), \ldots, y(x^{N}))^T$ denote the observed noisy samples at locations $\vec{x} = x^{1:N}$.  Given the data $(x,y)^{1:N}$ and the kernel $\cK$, the posterior of $\myf$ again forms a GP; in other words any collection $\myM_{x'_1}(\cdot), \ldots, \myM_{x'_k}(\cdot)$ is multivariate Gaussian with means $m(x'_i)$, covariances $v(x'_i,x'_j)$, and variances $v^2(x'_i) \equiv v(x'_i, x'_i)$, specified by the matrix equations  \cite[Sec.~2.7]{WilliamsRasmussenBook}: 
\begin{align}\label{eq:krig-mean}
  m(x) &=  \vec{k}(x)^T (\mathbf{K} + \mathbf{\Sigma})^{-1} \vec{y}; \\ \label{eq:krig-cov}
  v(x,x') & = {\cK}(x,x') - \vec{k}(x)^T (\mathbf{K} + \mathbf{\Sigma})^{-1} \vec{k}(x'), 
  \end{align}
with
$\vec{k}(x)  = (\cK(x^{1},x), \ldots, \cK(x^{N},x) )^T$, $\mathbf{\Sigma} := \diag( \sigma^2(x^{1}), \ldots, \sigma^2(x^{N}))$ and $\mathbf{K}$ the $N \times N$ positive definite matrix $\mathbf{K}_{i,j} := \cK(x^{i}, x^{j})$, $1 \le i,j \le N$. The diagonal structure of $\mathbf{\Sigma}$  is due to the assumption that independent simulations have uncorrelated noise, i.e.~use independent random numbers.
Note that the uncertainty, $v^2(x)$, associated
with the prediction at $x$ has no direct dependence on the simulation outputs $y^{1:N}$;
all response points contribute to the estimation
of the local error through their influence on the induced correlation matrix $\mathbf{K}$. Well-explored regions
will have lower $v^2(x)$, while regions with few or no observations (in particular regions beyond the training design) will have high $v^2(x)$.

\subsection{Training a Kriging Metamodel}\label{sec:train}
To fit a kriging metamodel requires specifying $\HH_{\cK}$, i.e.~fixing the kernel family and then estimating the respective kernel hyper-parameters, such as $s, \vec\theta$ in \eqref{eq:matern} that are plugged into \eqref{eq:krig-mean}-\eqref{eq:krig-cov}. The typical approach is to use a maximum likelihood approach which leads to solving a nonlinear optimization problem to find the MLEs $s^*, \vec\theta^*$ defining $\cK$. Alternatively, cross-validation techniques or EM-type algorithms are also available.

\red{While the choice of the kernel family is akin to the choice of orthonormal basis in LSMC, in our experience they play only a marginal role in performance of the overall pricing method. As a rule of thumb, we suggest to use either the Matern-5/2 kernel or the squared-exponential kernel. In both cases, the respective function spaces are smooth (twice-differentiable for Matern-5/2 family \eqref{eq:matern} and $C^\infty$ for the squared exponential kernel), and lead to similar fits. The performance is more sensitive to the kernel \emph{hyper-parameters}, which are typically estimated in a frequentist framework, but could also be inferred from a Bayesian prior, or even be guided by heuristics.}

\begin{remark}
  One may combine a kriging model with a classical least squares regression on a set of basis functions $\{ B_r \}$. This is achieved by taking $f(x) = t(x) + \tilde{f}(x)$ where $t(x) = \sum_r \beta_r B_r(x)$ is a trend term as in \eqref{eq:ols}, $\beta_r$'s are the trend coefficients to be estimated, and $\tilde{f}$ is the ``residual'' mean-zero Gaussian process. The Universal Kriging formulas, see \cite{Picheny10} or \cite[Sec.~2.7]{WilliamsRasmussenBook}, then allow simultaneous computation of the kriging surface and the OLS coefficients $\vec\beta$. For example, one may use the European option price, if one is explicitly available, as a basis function to capture some of the structure in $C(t,\cdot)$.
  \end{remark}

Inference on the kriging hyperparameters requires knowledge of the sampling noise $\sigma^2(x)$. Indeed, while it is possible to simultaneously train $\cK$ and infer a constant simulation variance $\sigma^2$ (the latter is known as the ``nugget'' in the machine learning community \cite{SantnerBook}), with state-dependent noise $\cK$ is not identifiable. Batching allows to circumvent this challenge by providing empirical estimates of $\sigma^2(x)$. For each site $x$, we thus sample $M$ independent replicates $y^{(1)}(x), \ldots, y^{(M)}(x)$ and estimate the conditional variance as
\begin{align}
  \label{eq:batch-sd} \widetilde{\sigma}^2(x) & := \frac{1}{M-1} \sum_{i=1}^M (y^{(i)}(x) -\bar{y}(x))^2, \quad\text{ where  } \quad \bar{y}(x) = \frac{1}{M} \sum_{i=1}^M y^{(i)}(x),
\end{align}
is the batch mean. We then use $\widetilde{\sigma}^2(x)$ as a proxy to the unknown $\sigma^2(x)$ to estimate $\cK$. One could further improve the estimation of $\sigma^2(\cdot)$ by fitting  an auxiliary kriging meta-model from $\widetilde{\sigma}^2$'s, cf.~\cite[Ch~5.6]{Kleijnen} and references therein. As shown in \cite[Sec 4.4.2]{GinsbourgerPicheny13}, after fixing $\cK$ we can then treat the $M$ samples at $x$ as the single design entry $(x,\bar{y}(x))$ with noise variance $\widetilde{\sigma}^2(x)/M$. The end result is that the batched dataset has just $N' := N/M$ rows $(x,\bar{y})^{1:N'}$ that are fed into the kriging meta-model.

For our examples below, we have used the  \texttt{R} package \textbf{DiceKriging} \cite{kmPackage-R}. The software takes the input locations $x^{1:N'}$, the corresponding simulation outputs $y^{1:N'}$ and the noise levels $\widetilde{\sigma}^2(x^{1:N'})$, as well as the kernel family (Matern-5/2 \eqref{eq:matern} by default) and returns a trained kriging model. One then can utilize a \texttt{predict} call to evaluate \eqref{eq:krig-mean}-\eqref{eq:krig-cov} at given $x'$s. The package can also implement universal kriging. Finally, in the cases where the design is not batched or the batch size $M$ is very small (below 20 or so), we resort to assuming homoscedastic noise level $\sigma^2$ that is estimated as part of training the kriging model (an option available in \textbf{DiceKriging}). \red{The advantage of such plug-and-play functionality is access to an independently tested, speed-optimized, debugged, and community-vetted code, that minimizes implementation overhead and future maintenance (while perhaps introducing some speed ineffiencies).}

\begin{remark}
  \red{The kriging framework includes   several well-developed generalizations, including treed GP's \cite{tgpPackage}, local GP's \cite{gramacy:apley:2013}, monotone GP's \cite{RiihimakiVehtari10}, and particle-based GP's \cite{GramacyPolson11} that
  can be substituted instead of the vanilla method described above, similar to replacing plain OLS with more sophisticated approaches. All the aforementioned extensions  can be used off-the-shelf through public \texttt{R} packages described in the above references. } 
\end{remark}

\subsection{Approximation Quality}
To quantify the accuracy of the obtained kriging estimator $\hat{C}(t,\cdot)$, we derive an empirical estimate of the loss function $L_{RMC}$. To do so, we integrate the posterior distributions $\myM_x(\cdot) \sim \mathcal{N}( m(x), v^2(x))$ vis-a-vis the surrogate means that are proxies for $C(t,\cdot)$ using \eqref{eq:loss}:
\begin{align}\notag
\ell_{RMC}(x; \ZZ) & := \int_\R |y-h(t,x)| 1_{\{m (x) < h(t,x) < y \bigcup \; y < h(t,x) < m(x) \}} \myM_x(dy) \\
\notag & = \int_0^\infty z \frac{1}{\sqrt{2 \pi}v(z)} \exp \left( - \frac{(m(z)-h(t,z))^2}{2 v^2(z)} \right) dz \\
\label{eq:emp-loss}
 & = {v(x)} \,\phi \!\left( \frac{-|m(x)-h(t,x)|}{v(x) } \right) - |m(x)-h(t,x)| \Phi\!\left(  \frac{-|m(x)-h(t,x)|}{v(x) } \right),
 \end{align}
where $\Phi, \phi$ are the standard Gaussian cumulative/probability density functions.
The quantity $\PP(\{m (x) < h(t,x) < C(t,x) \} \bigcup \{ C(t,x) < h(t,x) < m(x) \} \, \big| \, \GG)$ is precisely the Bayesian posterior probability of making the wrong exercise decision at $(t,x)$ based on the information from $\ZZ$, so lower $\ell_{RMC}(x; \ZZ)$ indicates better performance of the design $\ZZ$ in terms of stopping optimally. Integrating $\ell_{RMC}(\cdot)$ over $\XX$ then gives an estimate for $L_{RMC}(\hat{C})$:
\begin{align}\label{eq:hat-L}
  \widehat{L}_{RMC}(\hat{C}; \ZZ) := \int_{\XX} \ell_{RMC}(x; \ZZ) p(t,x|0,X_0)\, dx.
\end{align}
Practically, we replace the latter integral with a weighted sum over $\ZZ$, $\widehat{L}_{RMC}(\hat{C}; \ZZ) \simeq n^{-1} \cdot \sum_n \ell_{RMC}(x^n; \ZZ) p(t,x^n|0,X_0)$.

\subsection{Further RKHS Regression Methods}
From approximation theory perspective, kriging is an example of a regularized kernel regression. The general formulation  looks for the minimizer $\hat{f} \in \HH$ of the following penalized residual sum of squares problem
\begin{align}\label{eq:spline}
RSS(f, \la) = \sum_{n=1}^{N} \{y^{n} - \myf(x^{n})\}^2 + \la \| f \|^2_{\HH},
\end{align}
where $\lambda \ge 0$ is a chosen smoothing parameter and $\HH$ is a RKHS. The summation in \eqref{eq:spline} is a measure of closeness of $f$ to data, while the right-most term penalizes the fluctuations of $f$ to avoid over-fitting.
The representer theorem implies that the minimizer of \eqref{eq:spline} has an expansion in terms of the eigen-functions
\begin{equation}\label{eq:reg-regression}
\hat{f}(x) = \sum_{n=1}^{N} \alpha_n \cK(x,{x}^{n}),
\end{equation}
relating the prediction at $x$ to the kernel functions based at the design sites $x^{1:N}$; compare to \eqref{eq:krig-mean}.  In kriging, the function space is $\HH_{\cK}$, $\la=1/2$, and the corresponding norm $\| \cdot \|_{\HH}$ has a spectral decomposition in terms of differential operators \cite[Ch.~6.2]{WilliamsRasmussenBook}.

Another popular version of \eqref{eq:spline} are the smoothing (or thin-plate) splines that take $\cK_{TPS}(x, x') = \| x - x'\|^2 \log \| x - x' \|, $
where $\|\cdot\|$ denotes the Euclidean norm in  $\R^d$. In this case, the RKHS $\HH_{TPS}$ is the set of all twice continuously-differentiable functions \cite[Chapter 5]{FHT} and \begin{equation}\label{eq:2d-tpsconstraint}
 \| f \|^2_{\HH_{TPS}} = \int_{\R^d} \Bigl[ \sum_{i,j=1}^d \frac{\partial}{\partial x_i}\frac{ \partial}{\partial x_j} \myf( x)\Bigr] dx.
\end{equation}
This generalizes the one-dimensional penalty function $\| f\| = \int_\R \{f''(x)\}^2 dx$. Note that under $\la=\infty$ and \eqref{eq:2d-tpsconstraint} the optimization in \eqref{eq:spline} reduces to the traditional least squares linear fit $\hat{f}({x}) = \beta_0 + \beta_1 {x}$ since it introduces the constraint $f''({x}) = 0.$  A common parametrization for the smoothing parameter $\la$ is through the effective degrees of freedom statistic df$_\la$; one may also select $\la$ adaptively via cross-validation or MLE \cite[Chapter 5]{FHT}. 
The resulting optimization of \eqref{eq:spline} based on \eqref{eq:2d-tpsconstraint} gives a smooth $\mathcal{C}^2$ response surface which is called a thin-plate spline (TPS), and has the explicit form
\begin{equation}\label{eq:tps-solution}
\hat{\myf}(x) = \beta_0 + \sum_{j=1}^d \beta_j x_j + \sum_{n=1}^{N} \alpha_n \|x-x^{n}\|^2 \log \|x-x^{n}\|.
\end{equation}
See \cite{Kohler08spline} for implementation of RMC via  splines.

 A further RKHS approach, applied to RMC in \cite{TompaidisYang13} is furnished by the so-called  (Gaussian) radial basis functions. These are based on the already mentioned Gaussian kernel $\cK_{RBF}(x,x') = \exp(-\theta \| x-x'\|^2)$, but substitute the penalization in \eqref{eq:spline} with ridge regression, reducing the sum in \eqref{eq:reg-regression} to $N' \ll N$ terms by identifying/optimizing the ``prototype'' sites $x_{n'}$.

\section{Designs for Kriging RMC}\label{sec:designs}

Fixing the meta-modeling framework, we turn to the problem of generating the experimental design $\ZZ = x^{1:N}$. We work in the fixed-budget setting, with a pre-specified number of simulations $N$ to run.
\emph{Optimally} constructing the respective experimental design $\ZZ^{(N)}$ requires solving a $N$-dimensional optimization problem, which is generally intractable. Accordingly, in this section we discuss heuristics for generating near-optimal static designs. \red{Section \ref{sec:sequential}} then considers another work-around, namely sequential methods that utilize a divide-and-conquer approach.

\subsection{Space Filling Designs}
\red{A standard strategy in experimental design is to spread out the locations $x^{1:N}$ to extract as much information as possible from the samples $y^{1:N}$.  A \emph{space-filling} design attempts to distribute $x^n$'s evenly in the input space $\widetilde{\XX}$, here distinguished from the theoretical domain $\XX$ of $(X_t)$. While this approach does not directly target the objective \eqref{eq:loss}, assuming $\widetilde{\XX}$ is reasonably selected, it can still be much more efficient than traditional LSMC designs.} Examples of space-filling approaches include
maximum entropy designs, maximin designs, and maximum expected information designs \cite{Kleijnen}. Another choice are quasi Monte Carlo methods that use a deterministic space-filling sequence such as the Sobol for generating $x^{1:N}$. The simplest choice are gridded designs that pick $x^{1:N}$ on a pre-specified lattice, see Figure \ref{fig:interp} below. Quasi-random/low-discrepancy sequences have already been used in American option pricing but for other purposes; in \cite{BoyleTan13} for approximating one-step value in a stochastic mesh approach and in \cite{Chaudhary05} for generating the trajectories of $(X_t)$. Here we propose to use quasi-random sequences directly for constructing the simulation design; see also \cite{TompaidisYang14} for using a similar strategy in a stochastic control application.

One may also generate \emph{random} space-filling designs which is advantageous in our context of running multiple response surface models (indexed by $t$) by avoiding any grid-ghosting effects. One flexible procedure is Latin hypercube sampling (LHS) \cite{McBeCo:79}. In contrast to deterministic approaches that are generally only feasible in hyper-rectangular $\widetilde{\XX}$'s, LHS can be easily combined with an acceptance-rejection step to generate space-filling designs on any finite subspace.

Typically the theoretical domain $\XX$ of $(X_t)$ is unbounded (e.g.~$\RR^d_+$) and so the success of a space-filling design is driven by the judicious choice of an appropriate $\widetilde{\XX}$. This issue is akin to specifying the domain for a numerical PDE solver and requires structural knowledge. For example, for a Bermudan Put with strike $K$, we might take $\widetilde{\XX} = [0.6K, K]$, covering a wide swath of the in-the-money region, while eschewing out-of-the-money and very deep in-the-money scenarios. Ultimately, the problem setting determines the importance of this choice (compare the more straightforward setting of Section \ref{sec:heston2d} below vs.~that of Section \ref{sec:max-call} where finding a good $\widetilde{\XX}$ is much harder).

\subsection{Probabilistic Designs}\label{sec:prob-design}
The original solution of Longstaff and Schwartz~\cite{LS} was to construct the design $\ZZ$ using the conditional density of $X_t|X_0$, i.e.~to generate $N$ independent draws $x^{n} \sim p(t,\cdot|0,X_0)$, $n=1,\ldots,N$. This strategy has two advantages. First, it permits to implement the backward recursion for estimating $\mk{S}_{T-1}, \mk{S}_{T-2}, \ldots$ on a fixed global set of scenarios which requires just a single (albeit very large) simulation and hence saves simulation budget. Second, a probabilistic design automatically adapts to the dynamics of $(X_t)$, picking out the regions that $X_t$ is likely to visit and removing the aforementioned challenge of specifying $\widetilde{\XX}$.

Moreover, empirical sampling of $x^n$  reflects  the density of $X_t$ which helps to lower the weighted loss \eqref{eq:loss}. Assuming that the boundary $\partial\mk{S}_t$ is not in a region where $p(t,\cdot|0,X_0)$ is very low, this generates a well-adapted design, which partly explains why the experimental design of LSMC is not entirely without its merits. Compared to a space-filling design that is sensitive to $\widetilde{\XX}$, a probabilistic design is primarily driven by the initial condition $X_0$. This can be a boon as above, or a limitation; for example with OTM Put contracts, the vast majority of empirically sampled sites $x^n$ would be out-of-the-money, dramatically lowering the quality of an empirical $\ZZ$.
A good compromise is to take $x^n \sim X_t| \{X_0,h(t,X_t)>0\} $ i.e.~to condition on being in-the-money.

\subsection{Batched Designs}
As discussed, the design sites $x^{1:N}$ need not be distinct and the macro-design can be replicated. Such batched designs
offer several benefits in the context of meta-modeling. First, averaging of simulation outputs dramatically improves the statistical properties of the averaged $\bar{y}$ compared to the raw $y$'s.
In most cases once $M \gg 10$, the Gaussian assumption regarding the respective noise $\bar{\epsilon}$ becomes excellent. Second, batching raises the signal-to-noise ratio and hence simplifies response surface modeling. Training the kriging kernel $\cK$ with very noisy samples is difficult, as the likelihood function to be optimized possesses multiple local maxima. Third, as mentioned in Section \ref{sec:train}, batching allows empirical estimation of heteroscedastic sampling noise \red{$\sigma^2(x)$}. Algorithm \ref{algo:rmc} summarizes the steps for building a kriging-based RMC solver with a replicated design.

Batching also connects to the general bias-variance tradeoff in statistical estimation. Plain (pointwise) Monte Carlo offers a zero bias / high variance estimator of $f(x)$, which is therefore computationally expensive. A low-complexity model such as parametric  least squares \eqref{eq:ols}, offers a low variance/high bias estimator (since it requires constraining $\hat{f} \in \HH_R$). Batched designs coupled with kriging interpolate these two extremes by reducing bias through flexible approximation classes, and reducing variance through batching and modeling of spatial response correlation.

\begin{remark}
The batch sizes $M$ can be adaptive.  By choosing $M(x)$ appropriately one can set $\bar{\sigma}(x) \equiv Var( \bar{Y}(x)) = \widetilde{\sigma}^2(x)/M(x)$ to any desired level.  In particular, one could make all averaged observations homoscedastic with a pre-specified intrinsic noise $\bar{\sigma}$. The resulting regression model for $\bar{y}$ would then conform very closely to classical homoscedastic Gaussian assumptions regarding the distribution of simulation noise.
\end{remark}

\begin{algorithm}[ht]
\caption{Regression Monte Carlo for Optimal Stopping using Kriging \label{algo:rmc}}
\begin{algorithmic}[1]
\REQUIRE No.~of simulations $N$, no.~of replications $M$, no.~of time-steps $T$ 
\STATE $N' \leftarrow N/M$
\STATE Set $\widehat{\mk{S}}_T \leftarrow \XX$
\FOR{$t=T-1,T-2,\ldots,1$}
\STATE Generate a macro-design $\ZZ_t = \ZZ^{(N')}_t$
\STATE Use the stochastic sampler $Y(x) = H_t(X_\cdot)$ in \eqref{eq:H} with $M$ replications per $x \in \ZZ_t$
\STATE Batch the replications according to \eqref{eq:batch-sd} to obtain the averaged $(x,\bar{y})^{1:N'}$
\STATE Fit a kriging meta-model $\hat{C}(t,\cdot)$ based on $(x,\bar{y})^{1:N'}$
\STATE Obtain stopping set $\widehat{\mk{S}}_t$ from \eqref{eq:hatS-t}
\STATE (Or replace steps 4-8 with a sequential design sub-problem using Algorithm \ref{algorithm} below)
\ENDFOR
\STATE (Generate fresh out-of-sample $(X_0,y)^{1:{N}_{out}}$ and approximate $\hat{V}(0,X_0)$ using \eqref{eq:mc-average})
\RETURN Estimated stopping sets $\widehat{\mk{S}}_{1:T}$
\end{algorithmic}
\end{algorithm}

 As  the number of replications $M$ becomes very large, batching effectively eliminates all sampling noise. Consequently, one can substitute the empirical average $\bar{y}(x)$ from \eqref{eq:batch-sd} for the true $f(x)$, an approach first introduced in \cite{VanBeersKleijnen03}. It now remains only to interpolate these responses, reducing the meta-modeling problem to analysis of \emph{deterministic} experiments. There is a number of highly efficient interpolating meta-models, including classical deterministic kriging (which takes $\Sigma \equiv \mathbf{0}$ in \eqref{eq:krig-mean}-\eqref{eq:krig-cov}), and natural cubic splines. Assuming smooth underlying response,  deterministic meta-models often enjoy very fast convergence. Application of interpolators offers a new perspective (which among others is nicer to visualize) on the RMC meta-modeling problem. To sum up, batching offers a tunable spectrum of methods that blend smoothing and interpolation of $\myf(\cdot)$.

\begin{figure}[ht]
  \centering
 \includegraphics[width=0.65\textwidth,height=2.3in,trim=0.25in 0.15in 0.15in 0.15in]{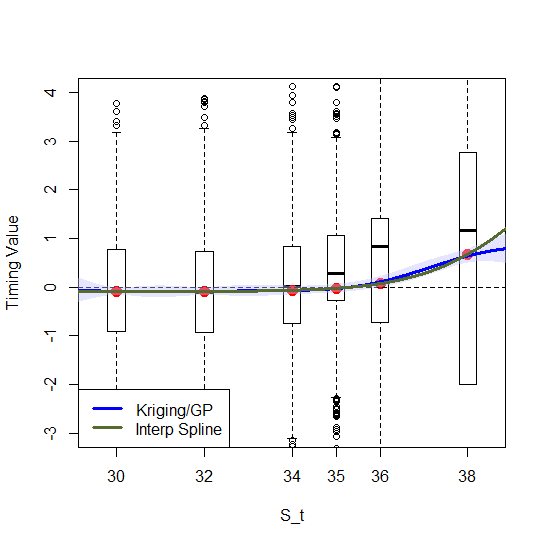}
  \caption{Experimental design and estimated timing value $\hat{T}(t,\cdot)$ using (i) deterministic kriging and  (ii) a cubic spline interpolator. We use the manually chosen macro design $\ZZ' = x^{1:N'} = \{30,32,34,35,36,38\}$, with $M=1600$ replications at each site.  The boxplots summarize the resulting distribution of $y^{(m)}(x^n)$'s, $m=1,\ldots,M$. The dots indicate the batch means $\bar{y}$'s which are interpolated exactly by the two meta-models. The underlying Bermudan Put is as in Section \ref{sec:1d}. \label{fig:interp}}
\end{figure}

Figure \ref{fig:interp} illustrates the above idea for a 1-D  Bermudan Put in a GBM model, see Section \ref{sec:1d}. We construct a design $\ZZ$ of macro size $9600 = 1600 \cdot 6$ which uses just six distinct sites $\ZZ' = x^{1:N'}$, and $M=1600$ replications at each site. We then interpolate the obtained values $\bar{y}(x^{1:6})$ using (i) a deterministic kriging model; and a (ii) natural cubic spline. We observe that the resulting fit for $\hat{T}(t,\cdot)$ shown in the Figure looks excellent and yields an accurate approximation of the exercise boundary.

\subsection{Illustration}\label{sec:1d}
To illustrate different simulation designs we revisit the classical example of a 1-D  Bermudan Put option in a Geometric Brownian motion model. \red{More extensive experiments are given in Section \ref{sec:examples}.}  The log-normal $X$-dynamics are
\begin{align}\label{eq:gbm}
X_{t +\Delta t} = X_t \exp \left({ (r-\delta -\frac{1}{2} \sigma^2) \Delta t + \sigma  \Delta W_t}\right), \qquad \Delta W_t \sim \mathcal{N}(0, \Delta t),
\end{align}
and the Put payoff is $h(t,x) = e^{-rt}(K-x)_+$. The option matures at $T=1$, and we assume 25 exercise opportunities spaced out evenly with $\Delta t= 0.04$. This means that with a slight abuse of notation, the RMC algorithm is executed over $t=T, T-\Delta t, \ldots, \Delta t, 0$.  The rest of the parameter values are interest rate $r=0.06$, dividend yield $\delta=0$, volatility $\sigma = 0.2$, and at-the-money strike $K=X_0 = 40$. In this case $V(0,X_0) = 2.314$, cf.~\cite{LS}. In one dimension, the stopping region for the Bermudan Put is an interval $[0,\underline{s}(t)]$ so there is a  unique exercise boundary $\underline{s}(t) = \partial \mk{S}_t$.

We implement Algorithm \ref{algo:rmc} using a batched LHS design $\ZZ$ with $N=3000, M=100$, so at each step there are $N'=N/M=30$ distinct simulation sites $x^{1:N'}$. The domain for the experimental designs is the in-the-money region $\widetilde{\XX} = [25,40]$. To focus on the effect of various $\ZZ$'s, we \emph{freeze} the kriging kernel $\cK$ across all time steps $t$ as a Matern-5/2 type \eqref{eq:matern} with hyperparameters $s=1, \theta = 4$, which were close to the MLE estimates obtained. This choice is maintained for the rest of this section, as well as for the following Figures \ref{fig:T-over-time}-\ref{fig:sur-learning}. \red{Our experiments (see in particular Table \ref{tbl:kernel}) imply that the algorithm is insensitive to the kernel family or the specific software/approach used for training the kriging kernels.}

\begin{figure}[ht]
  \centering
 \hspace*{-0.15in}\begin{tabular}{ccc}
 \includegraphics[width=0.32\textwidth,height=2.4in,trim=0in 0.1in 0.0in 0.1in]{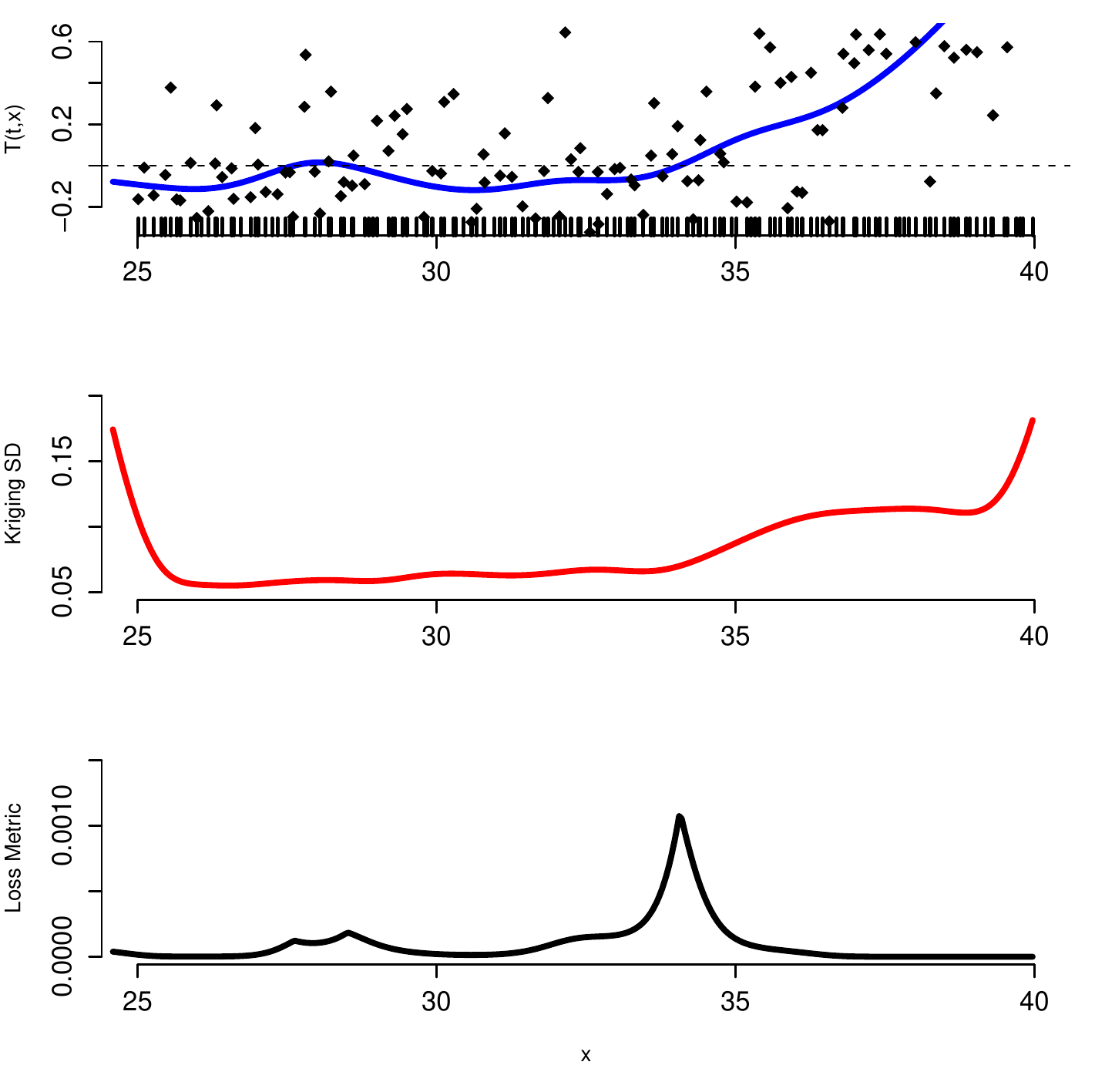} &
  \includegraphics[width=0.32\textwidth,height=2.4in, trim=0in 0.1in 0.0in 0.1in]{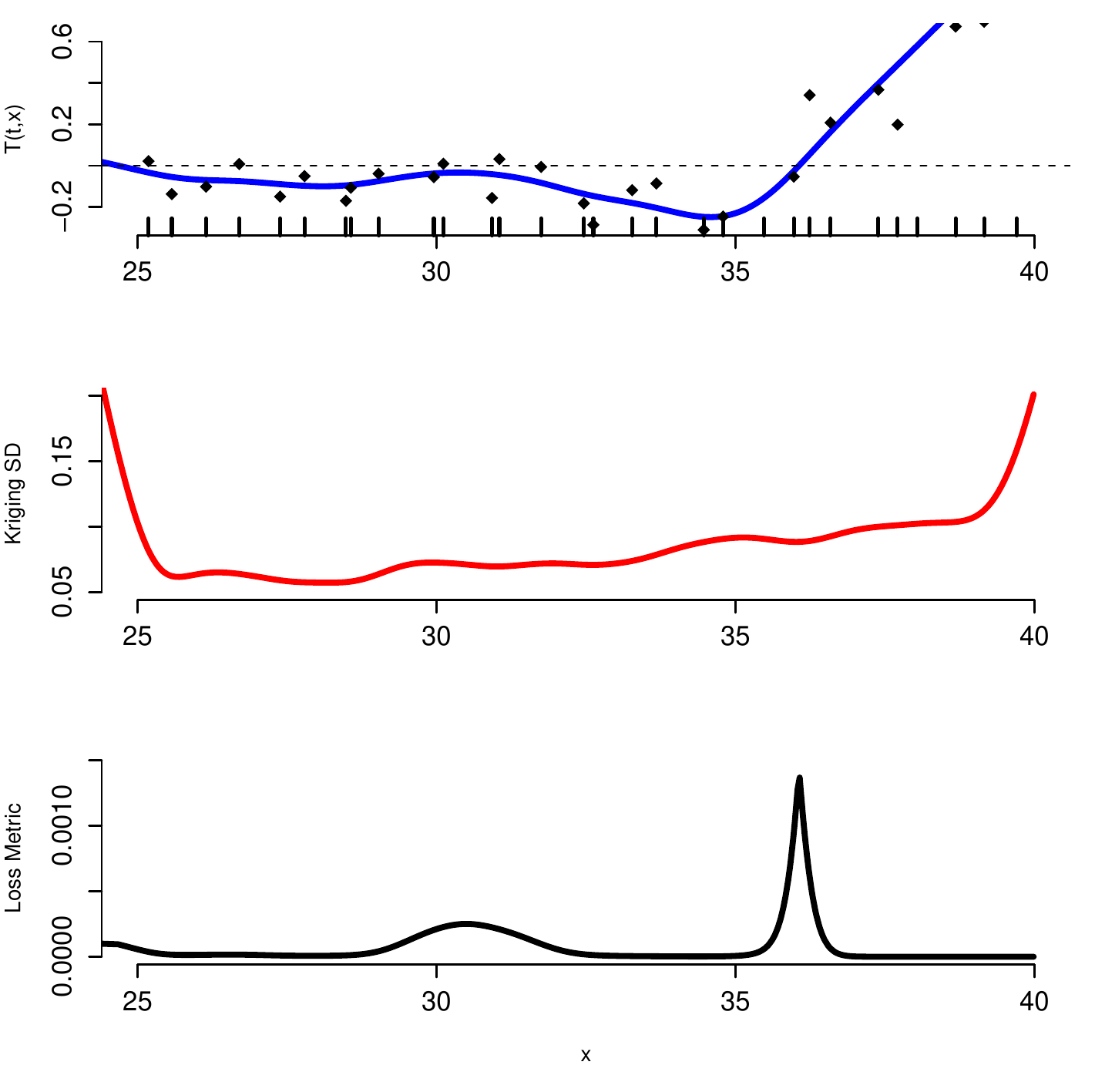} &
  \includegraphics[width=0.32\textwidth,height=2.4in, trim=0in 0.1in 0.0in 0.1in]{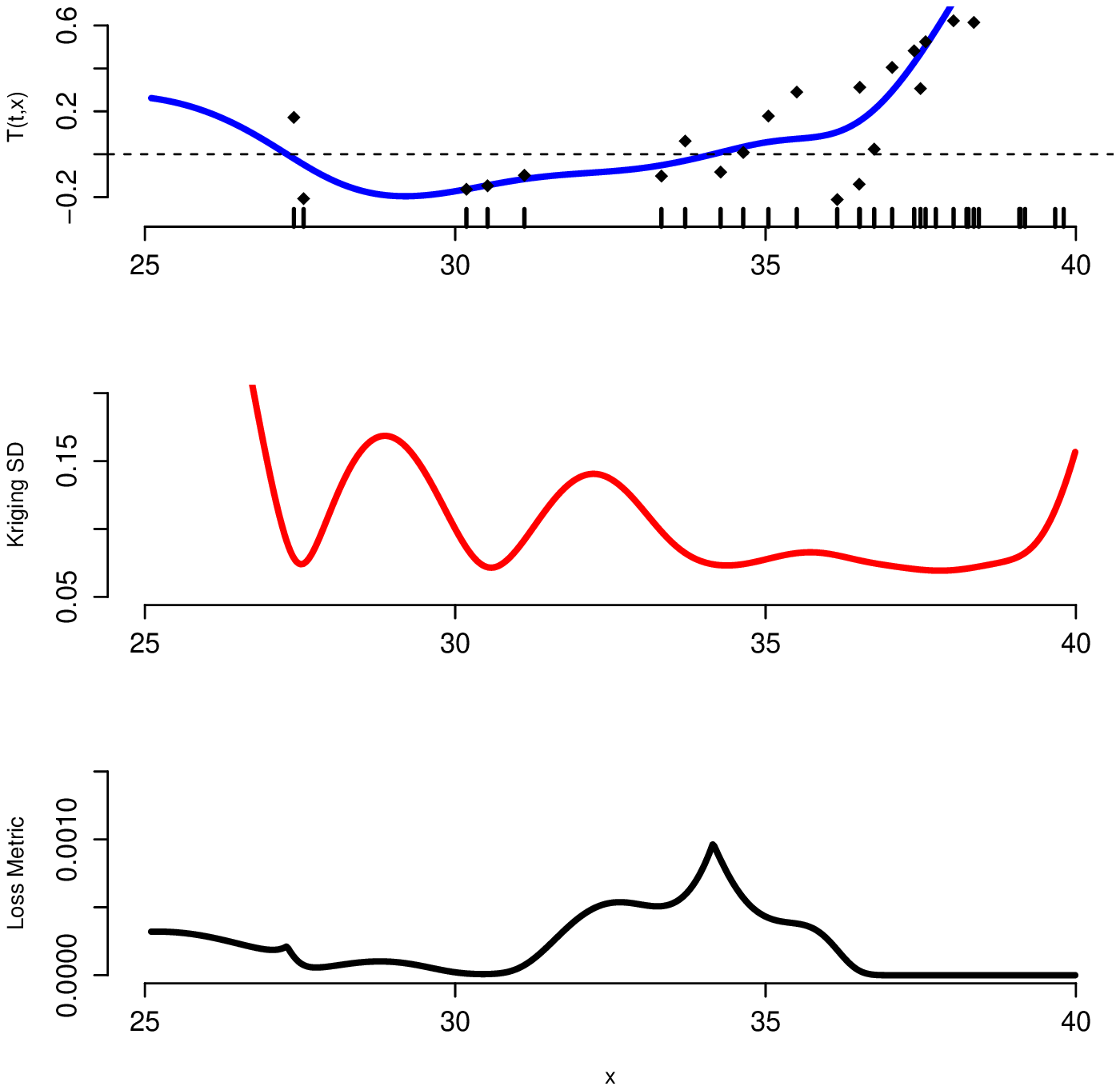} \\
  LHS $M=20, N'=150$ & LHS $M=100, N'=30$ & Emp $M=100, N'=30$ \end{tabular}
  \caption{Three different designs for fitting a kriging metamodel of the continuation value for the 1-D Bermudan Put ($t=0.6, T=1$). \emph{Top} panels show the fitted $\hat{T}(t,\cdot)$ as well as the distinct design sites $x^{1:N'}$. \emph{Middle} panels plot the corresponding surrogate standard deviation $v(x)$. \emph{Bottom} panels display the loss metric $\ell(x; \ZZ)$ from \eqref{eq:emp-loss}. \label{fig:designs} }
\end{figure}

Figure \ref{fig:designs} illustrates the effect of various experimental designs on the kriging meta-models.
We compare three different batched designs with a fixed overall size $N = |\ZZ| = 3000$: (a) an LHS design with small batches $M=20$; (b) an LHS design with a large $M=100$, and (c) \red{a probabilistic density-based design also with $M=100$; recall that the latter generates $N/M=30$ paths of $(X^x_t)$ and then creates $M$ sub-simulations initiating at each $x^k_t$.}
The resulting $(x_t, y_t)^{1:N'}$ at $t=0.6$ are shown in the top panels of Figure \ref{fig:designs}, along with the obtained estimates $\hat{T}(t,\cdot)$.
The middle row of Figure \ref{fig:designs} shows the resulting surrogate standard deviations $v(\cdot)$. The shape of $x \mapsto v(x)$ is driven by the local density of the respective $\ZZ$, as well as by the simulation noise $\sigma^2(x)$. Here, $\sigma^2(x)$ is highly heteroscedastic, being much larger near at-the-money $x \simeq 40$ than deep ITM. This is because in-the-money one is likely to stop soon and so the variance of $H_t(X^x_\cdot)$ is lower. For LHS designs, the roughly uniform density of $\ZZ$ leads to the posterior variance being proportional to the simulation noise, $v^2(x) \propto \sigma^2(x)$; on the other hand the empirical design reflects the higher density of $X_t$ closer to $X_0=40$, so that $v(x)$ is smallest there. Moreover, because there are very few design sites for $x<32$ (just 5 in Figure \ref{fig:designs}), the corresponding surrogate variance has the distinctive hill-and-valley shape. In all cases, the surrogate variance explodes along the edges of $\ZZ$, reflecting lack of information about the response there.

The bottom panels of Figure \ref{fig:designs} show the local loss metric $\ell_{RMC}(x; \ZZ)$ for the corresponding designs. We stress that $\ell_{RMC}(\cdot)$ is a function of the sampled $y^{1:N}$, and hence will vary across algorithm runs. Intuitively $\ell_{RMC}(\cdot)$ is driven by the respective surrogate variance $v^2(x)$, weighted in terms of the distance $|m(x) - h(t,x)|$ to the stopping boundary. Consequently, large $v^2(x)$ is fine for regions far from the exercise boundary. Overall, we can make the following observations:
\renewcommand{\theenumi}{\roman{enumi}}
 \begin{enumerate}
\item Replicating simulations does not materially impact surrogate variance at any given $x' \notin \ZZ$, and hence makes little effect on $\ell_{RMC}(x')$. Consequently, there is minimal loss of fidelity relative to using a lot of distinct design sites in $\ZZ$.
\item The modeled response surfaces tend to be smooth. Consequently, the kernel $\cK$ should enforce long-range spatial dependence (sufficiently large lengthscale) to make sure that $\hat{T}(t,\cdot)$ is likewise smooth. Undesirable fluctuations in $\hat{T}(t,\cdot)$ can generate spurious stopping/continuation regions, see e.g.~the left-most panel in Figure \ref{fig:designs} around $x=28$. 
\item Probabilistic designs tend to increase $\widehat{L}_{RMC}(\hat{C})$ because they fail to place enough design sites deep ITM, leading to extreme posterior uncertainty $v^2(x)$ there, cf.~the right panel in the Figure.
\item Due to very low signal-to-noise ratio that is common in financial applications, replication of simulations aids in seeing the ``shape'' of $C(t,\cdot)$ and hence simplifies the meta-modeling task, see point (ii).
\end{enumerate}

To give a sense regarding the aggregate quality of the meta-models,
Figure \ref{fig:T-over-time} shows the estimated $\hat{T}(t,\cdot)$ at three different steps $t$ as the RMC implements backward recursion. The overall shape of $\hat{T}(t,\cdot)$ remains essentially the same, being strongly positive close to the strike $K=40$, crossing zero at $\underline{s}$ and remaining slightly negative deep ITM. As $t$ decreases, the exercise boundary $\underline{s}_t$ also moves lower. The Figure shows how the cross-sectional borrowing of information (modeled by the GP correlation kernel $\cK$) reduces the surrogate variance $v^2(x)$ compared to the raw $\widetilde{\sigma}^2(x)$ (cf.~the corresponding 95\% CIs). We also observe that there is some undesirable ``wiggling'' of $\hat{T}(t,\cdot)$, which can generate spurious exercise boundaries such as in the extreme left of the $t=0.2$ panel in Figure \ref{fig:T-over-time}. However, with an LHS design this effect is largely mitigated compared to the performance of the empirical design in Figure \ref{fig:designs}. This confirms the importance of spacing out the design $\ZZ$.  Indeed, for the LHS designs the phenomenon of spurious exercising only arises for $t \simeq 0$ and $x$ very small, whereby the event $\{ X_t < 30 \}$ that would lead to the wrong exercise strategy has negligible probability of occurring.

\begin{figure}[ht]
  \centering
  \hspace*{-0.18in}
  \begin{tabular}{ccc}
 \includegraphics[width=0.32\textwidth,height=2.2in]{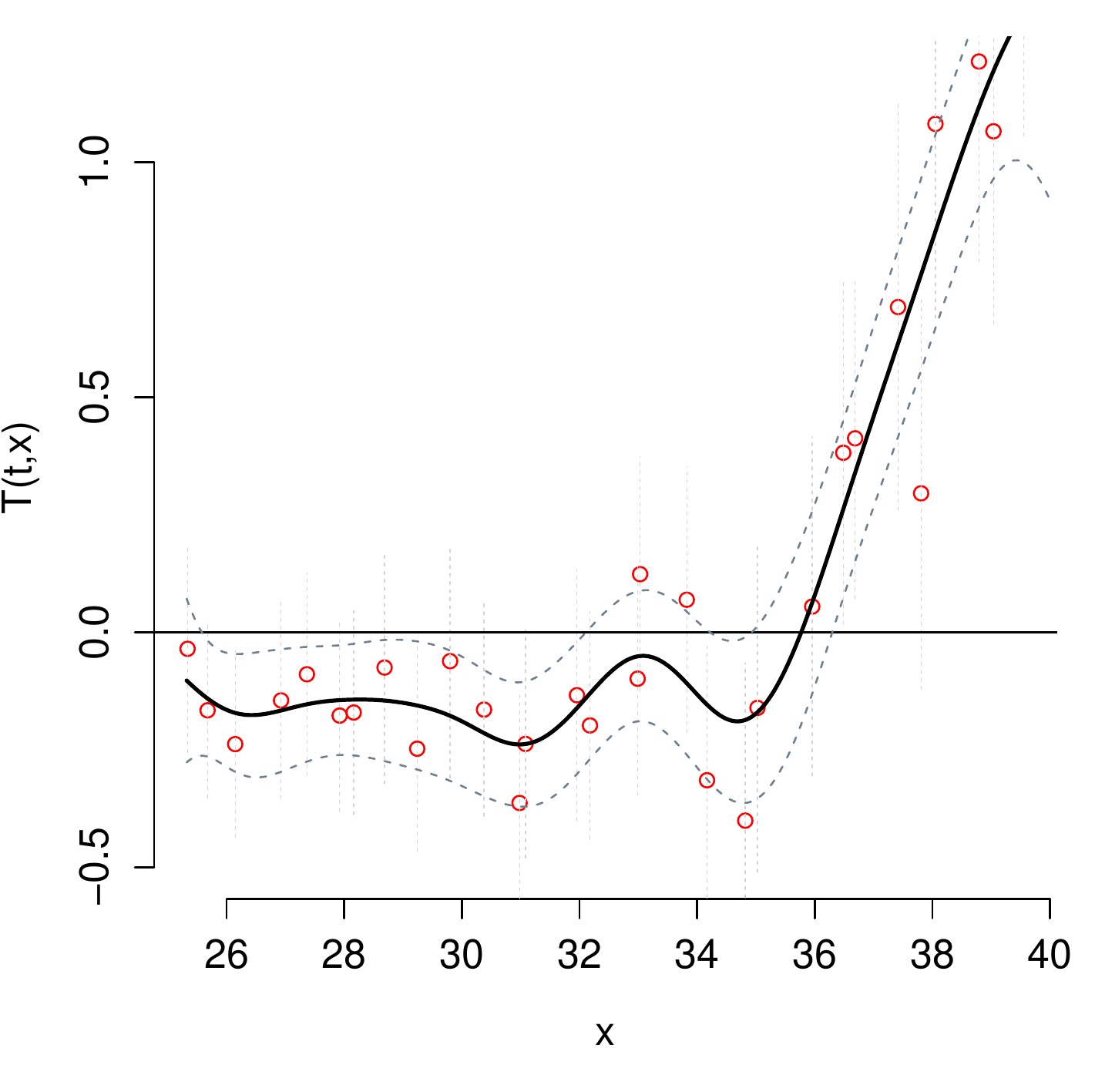} &
 \includegraphics[width=0.32\textwidth,height=2.2in]{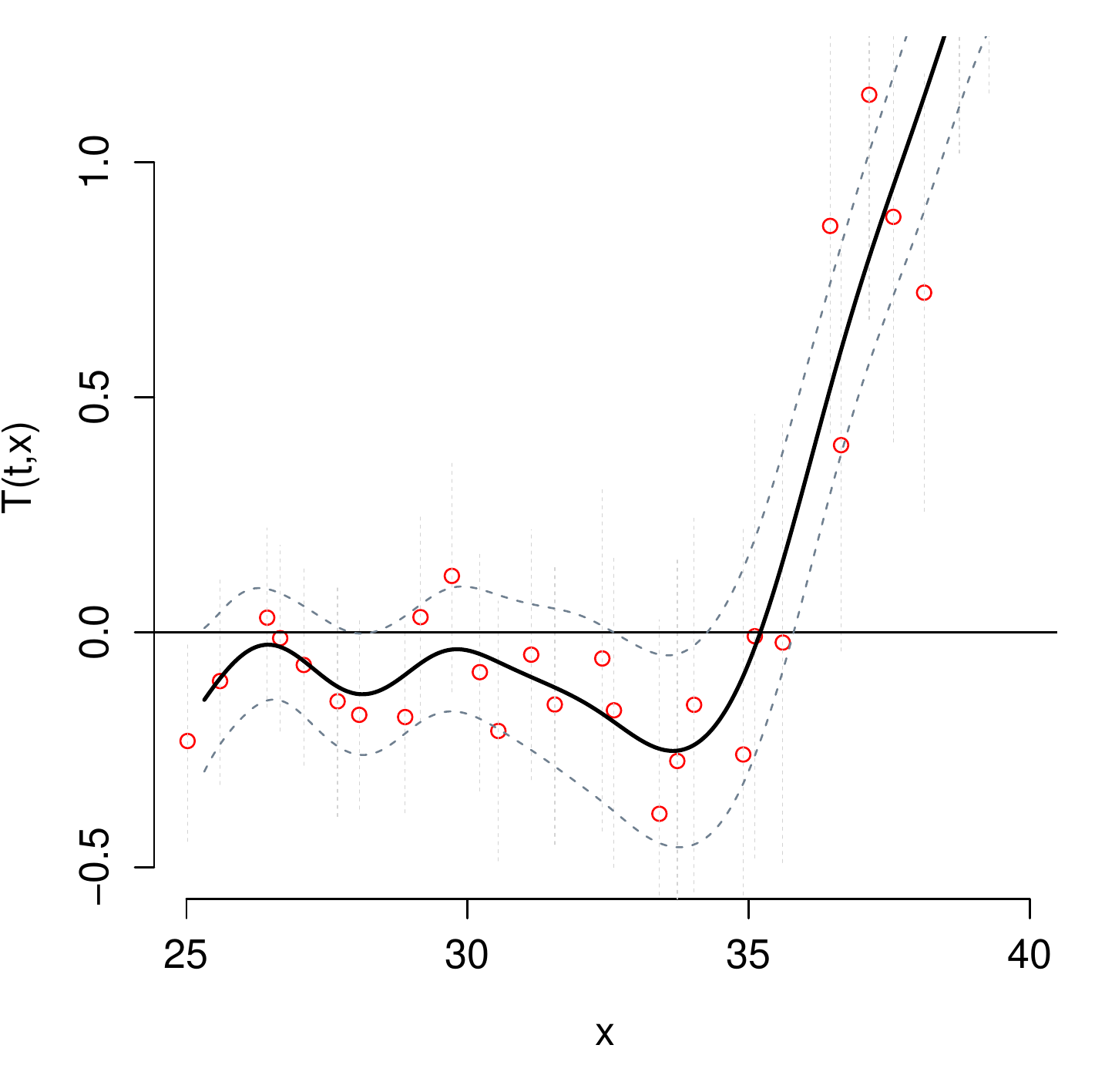} &
 \includegraphics[width=0.32\textwidth,height=2.2in]{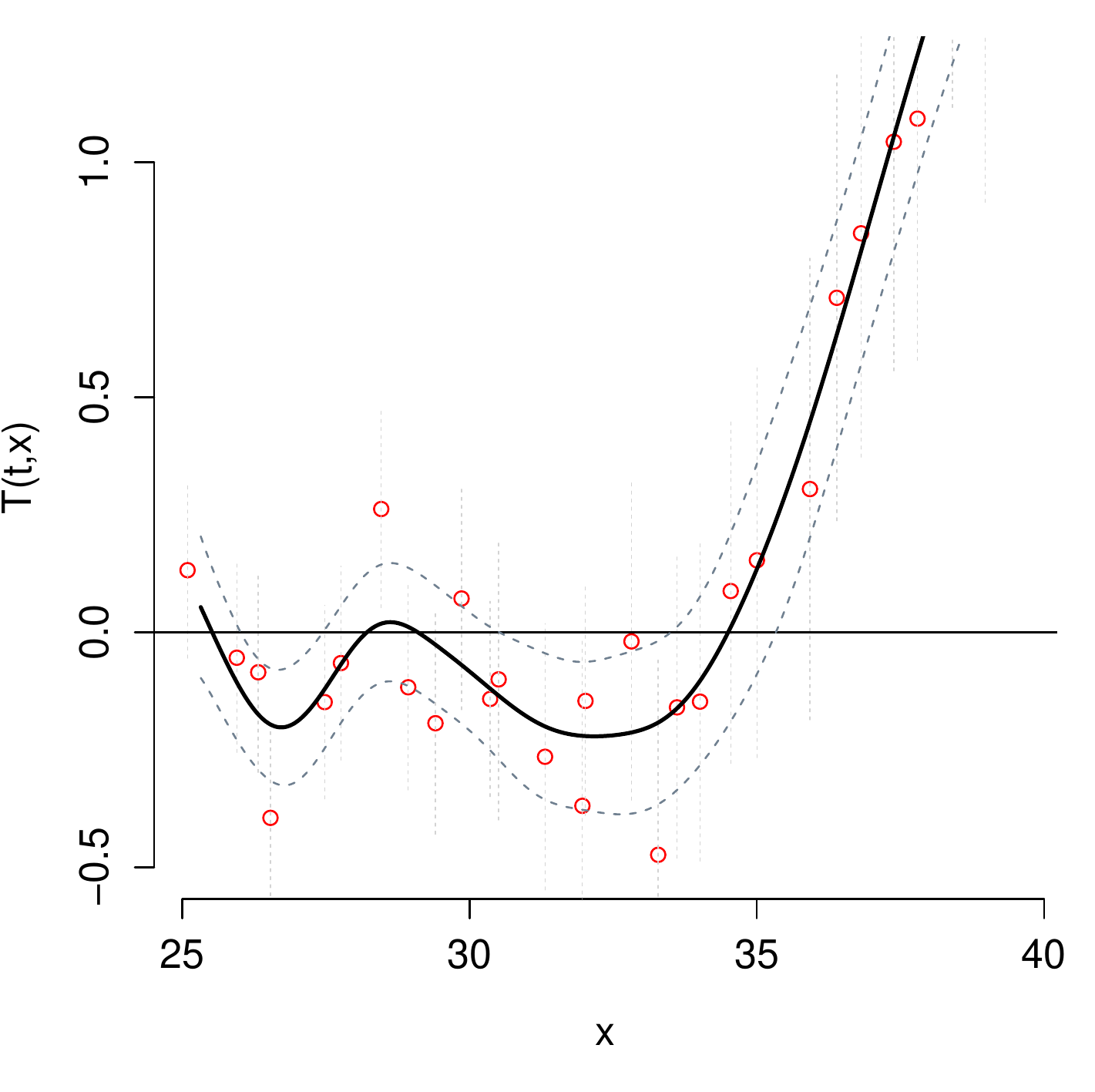} \\
 $t=0.6$ & $t=0.4$ & $t=0.2$ \end{tabular}
  \caption{Evolution of the estimated $\hat{T}(t,\cdot)$ over the backward induction steps. We show the snapshots at $t=0.6, 0.4, 0.2$. LHS designs $\ZZ$ of size $N=3000$ with $M=100$ replications. The vertical ``error'' bars indicate the 95\% quantiles of the simulation batch $y^{(i)}, i=1,\ldots,M$ at $x^n$, $n=1,\ldots, N'$ (based on \eqref{eq:batch-sd}); the dots show the batch means $\bar{y}$, while the dotted lines indicate the 95\% credibility interval (CI) of the kriging meta-model $\myM_{x}(\cdot)$. \label{fig:T-over-time} }
\end{figure}

\section{Sequential Designs}\label{sec:sequential}
In this section we discuss adaptive designs that are generated on the fly as the meta-model learns $C(t,\cdot)$.
Sequential design captures the intuition of focusing the sampling efforts around the exercise boundary $\partial {\mk{S}}_t$, cf.~the loss function \eqref{eq:loss}. This makes learning the response intrinsic to constructing the experimental design. In particular, Bayesian procedures embed sequential design within a  dynamic programming framework that is naturally married to statistical learning. Algorithmically,
sequential design is implemented at each time step $t$ by introducing an additional loop over $k=N_0,\ldots,N'$ that grows the experimental designs $\ZZ^{(k)}$ now indexed by their size $k$ ($k$ is the number of distinct sites, with batching possibly applied in the inner loop as before). As the designs are augmented, the corresponding surrogate surfaces $\myM^{(k)}_\cdot$ and stopping regions $\widehat{\mk{S}}^{(k)}_t$ are re-estimated and refined. The final result is a sequence of adaptive designs $\ZZ^{(N')}_{T-1}, \ZZ^{(N')}_{T-2},\ldots$ and corresponding exercise policy.

\subsection{Augmenting the Experimental Design}
Fixing a time step $t$, the sequential design loop is initialized with $\ZZ^{(N_0)}$, generated for example using LHS. New design sites $x^{N_0+1}, x^{N_0+2},\ldots$ are then iteratively added by greedily optimizing an acquisition function $E(x)$ that quantifies information gains via Expected Improvement (EI) scores. The aim of EI scores is to identify locations $x$ which are most promising in terms of lowering the global loss function $L_{RMC}(\hat{C})$ from \eqref{eq:loss}. In our context the EI scores $E_k(x)$ are based on the posterior distributions $\myM^{(k)}_x(\cdot)$ which summarize information learned so far about $C(t,\cdot)$. The seminal works of \cite{cohn:1996,Mackay92}
showed that a good heuristic to maximize learning rates is to  sample at sites that have high surrogate variance or high expected surrogate variance reduction, respectively. Because we target \eqref{eq:loss}, a second objective is to preferentially explore regions close to the contour $\{ \hat{C}(t,x) = h(t,x)\}$. This is achieved by blending the distance to the contour with the above variance metrics, in analogue to the Efficient Global Optimization approach \cite{JonesSchonlauWelch98} in simulation optimization.

A greedy strategy augments $\ZZ^{(k)}$ with the location that maximizes the acquisition function:
\begin{align}\label{eq:best-1}
  x^{k+1} = \argsup_{x \in \XX } E_k(x).
\end{align}
Two major concerns for implementing \eqref{eq:best-1} are (i) the computational cost of optimizing $x^{k+1}$ over $\XX$ and (ii) the danger of myopic strategies. For the first aspect, we note that \eqref{eq:best-1} introduces a whole new optimization sub-problem just to augment the design. This can generate substantial overhead that deteriorates the running time of the entire RMC. Consequently, approximate optimality is often a pragmatic compromise to maximize performance. For the second aspect, myopic nature of \eqref{eq:best-1} might lead to undesirable concentration of the design $\ZZ$ interfering with the convergence of $\hat{C}^{(k)}(t,\cdot) \to C(t,\cdot)$ as $k$ grows. It is well known that many greedy sequential schemes can get trapped sampling in some subregion of $\XX$, generating poor estimates elsewhere. For example, if there are multiple exercise boundaries, the EI metric might over-prefer established  boundaries, and risk missing other stopping regions that were not yet found. A common way to circumvent this concern is to randomize \eqref{eq:best-1}, which ensures that $\ZZ$ grows dense uniformly on $\tilde\XX$. In the examples below we follow \cite{GL13} and replace $\argsup_{ x \in \XX }$ in \eqref{eq:best-1} with $\arg\max_{x \in \TT }$ where $\TT$ is a finite \emph{candidate set}, generated using LHS over some domain $\tilde\XX$ again. LHS candidates ensure that potential new $x^{k+1}$ locations are representative and well spaced out over $\XX$.

Kriging meta-models are especially well-suited for sequential design thanks to availability of simple \emph{updating formulas} that allow to efficiently assimilate new data points into an existing fit. If a new sample $(x^{k+1},y^{k+1})$ is added to an existing design $x^{1:k} \equiv \ZZ^{(k)}$ and keeping the kernel $\cK$ fixed, the surrogate mean and variance at location $x$ are updated via
\begin{align}\label{eq:update-mean}
 m^{(k+1)}(x) &=  m^{(k)}(x) + \lambda(x, x^{k+1}; x^{1:k}) (y^{k+1} - m^{(k)}(x^{k+1})); \\ \label{eq:update-var}
  v^{(k+1)}(x)^2 & = v^{(k)}(x)^2 - \lambda(x, x^{k+1}; x^{1:k})^2 [\sigma^2(x^{k+1}) + v^{(k)}(x^{k+1})^2],
\end{align}
where $\lambda(x,x^{k+1}; x^{1:k}) > 0 $ is a weight function (that is available analytically) specifying the influence of the new $x^{k+1}$-sample on $x$, conditional on existing design locations $x^{1:k}$. Note that \eqref{eq:update-mean} and \eqref{eq:update-var} only require the knowledge of the latest surrogate mean/variance and $x^{1:k}$; previous simulation outputs $y^{1:k}$ do not need to be stored. Moreover, the updated surrogate variance $v^{(k+1)}(x)^2$ is a deterministic function of $x^{k+1}$ which is independent of $y^{k+1}$. In particular, the local reduction in surrogate \emph{standard deviation} at $x^{k+1}$ is proportional to the current $v^{(k)}(x^{k+1})$~\cite{GinsbourgerEmery14}:
\begin{align}\label{eq:krig-update}
  {v^{(k)}(x^{k+1}) - v^{(k+1)}(x^{k+1})} =  {v^{(k)}(x^{k+1})} \cdot \Bigl[1- \frac{ \sigma(x^{k+1}) }{ \sqrt{ \sigma^2(x^{k+1})+ v^{(k)}(x^{k+1})^2} } \Bigr].
\end{align}

\begin{remark}
  In principle, the hyperparameters of the kernel $\cK^{(k)}$ ought to be re-estimated as more data is collected, which makes updating $\hat{C}(t,\cdot)$ non-sequential. However, since we expect the estimated $\cK^{(k)}$ to converge as $k$ grows, it is a feasible strategy to keep $\cK$ frozen across sequential design iterations, allowing the use of \eqref{eq:update-mean}-\eqref{eq:update-var}.
\end{remark}

Algorithm \ref{algorithm} summarizes sequential design for learning $C(t,\cdot)$ at a given time-step $t$; see also Algorithm \ref{algo:rmc} for the overall kriging-based RMC approach to optimal stopping.

\begin{algorithm}[H]
\caption{Sequential design for \eqref{eq:loss} using stochastic kriging \label{algorithm}}
\begin{algorithmic}[1]
\REQUIRE Initial design size $N_0$, final size $N$, acquisition function $E(\cdot)$ 
\STATE Generate initial design $\ZZ^{(N_0)} := x^{1:N_0}$ of size $N_0$ using LHS over $\tilde\XX$
\STATE Sample $y^{1:N_0}$ and initialize the response surface model
\STATE Construct the stopping set $\mk{S}^{(N_0)}(\cdot)$ using \eqref{eq:hatS-t}
\STATE $k \leftarrow N_0$
\WHILE{$k < N$}
\STATE Generate a new candidate set $\TT^{(k)}$ of size $D$ using LHS
\STATE Compute the expected improvement $E_{k}(x)$ for each $x \in \TT$
\STATE Pick a new location $x^{k+1} = \arg\max_{x\in \TT^{(k)}} E_k(x)$ and sample the corresponding $y^{k+1}$
\STATE (Optional) Re-estimate the kriging kernel $\cK$
\STATE Update the surrogate surface using \eqref{eq:update-mean}-\eqref{eq:update-var}
\STATE Update the stopping set $\mk{S}^{(k+1)}$ using \eqref{eq:hatS-t}
\STATE Save the overall grid $\ZZ^{(k+1)} \leftarrow \ZZ^{(k)} \cup x^{k+1}$
\STATE k $\leftarrow$ k+1
\ENDWHILE
\RETURN Estimated stopping set $\mk{S}^{(N)}$ and design $\ZZ^{(N)}$
\end{algorithmic}
\end{algorithm}

\begin{remark}
  The ability to quickly update the surrogate as simulation outputs are collected lends kriging-models to ``online'' and parallel implementations. This can be useful even without going through a full sequential design framework. For example, one can pick an initial budget $N$, implement RMC with $N$ simulations, and if the results are not sufficiently accurate, add \emph{more} simulations without having to completely restart from scratch. Similarly, batching simulations is convenient for parallelizing using GPU technology.
\end{remark}

\subsection{Acquisition Functions}
In this section we propose several acquisition functions $E(x)$ to guide the sequential design sub-problem \eqref{eq:best-1}. Throughout we are cognizant of the loss function \eqref{eq:loss} that is the main criterion for learning $C(t,\cdot)$.

One possible proposal for an EI metric is to sample at locations that have a high (weighted) local loss $\ell^{(k)}(x) \equiv \ell_{RMC}(x; \mathcal{D}^{(k)})$ defined in \eqref{eq:emp-loss}, i.e.~
\begin{align}\label{eq:ei-zc}
E^{ZC}_k(x) & := \ell^{(k)}(x) \cdot p(t,x|0,X_0).
\end{align}
The superscript ZC stands for zero-contour, since intuitively $E^{ZC}(x)$ measures the weighted distance between $x$ and exercise boundary, see~\cite{GL13}. $E^{ZC}$ targets regions with high $v^{(k)}(x)$ or close to the exercise boundary, $|m^{(k)}(x) - h(t,x)| \simeq 0$.

A more refined version, dubbed ZC-SUR, attempts to identify regions where $\ell^{(k)}(x)$ can be quickly \emph{lowered} through additional sampling. To do so, ZC-SUR incorporates the expected difference $\EE[ \ell(x; \ZZ^{(k)}) - \ell(x; \ZZ^{(k)} \cup x) ] \ge 0$ which can be evaluated using the updating formulas \eqref{eq:update-mean}-\eqref{eq:update-var}. This is similar in spirit to the stepwise uncertainty reduction (SUR) criterion in stochastic optimization~\cite{Picheny10}. Plugging-in \eqref{eq:loss} we obtain
\begin{align}\label{eq:ei-sur}
  EI^{ZC-SUR}_k(x) & := p(t,x|0,X_0) \cdot \Biggl\{ {v^{(k)}(x)} \,\phi \left( \frac{-d^{(k)}(x)}{{v^{(k)}(x)} } \right)  - {v^{(k+1)}(x)} \,\phi \left( \frac{-d^{(k)}(x)}{{v^{(k+1)}(x)} } \right)
   \\ \notag & \quad - d^{(k)}(x) \left[ \Phi\Bigl(  \frac{-d^{(k)}(x)}{{v^{(k)}(x)} } \Bigr) - \Phi\Bigl(  \frac{-d^{(k)}(x)}{{v^{(k+1)}(x)} } \Bigr) \right] \Biggr\}, \quad d^{(k)}(x) := |m^{(k)}(x)-h(t,x)|.
\end{align}

\begin{figure}[ht]
  \hspace*{-0.18in}
  \begin{tabular}{cc}
 \includegraphics[width=0.47\textwidth,height=2.5in,trim=0.1in 0.1in 0.1in 0.1in]{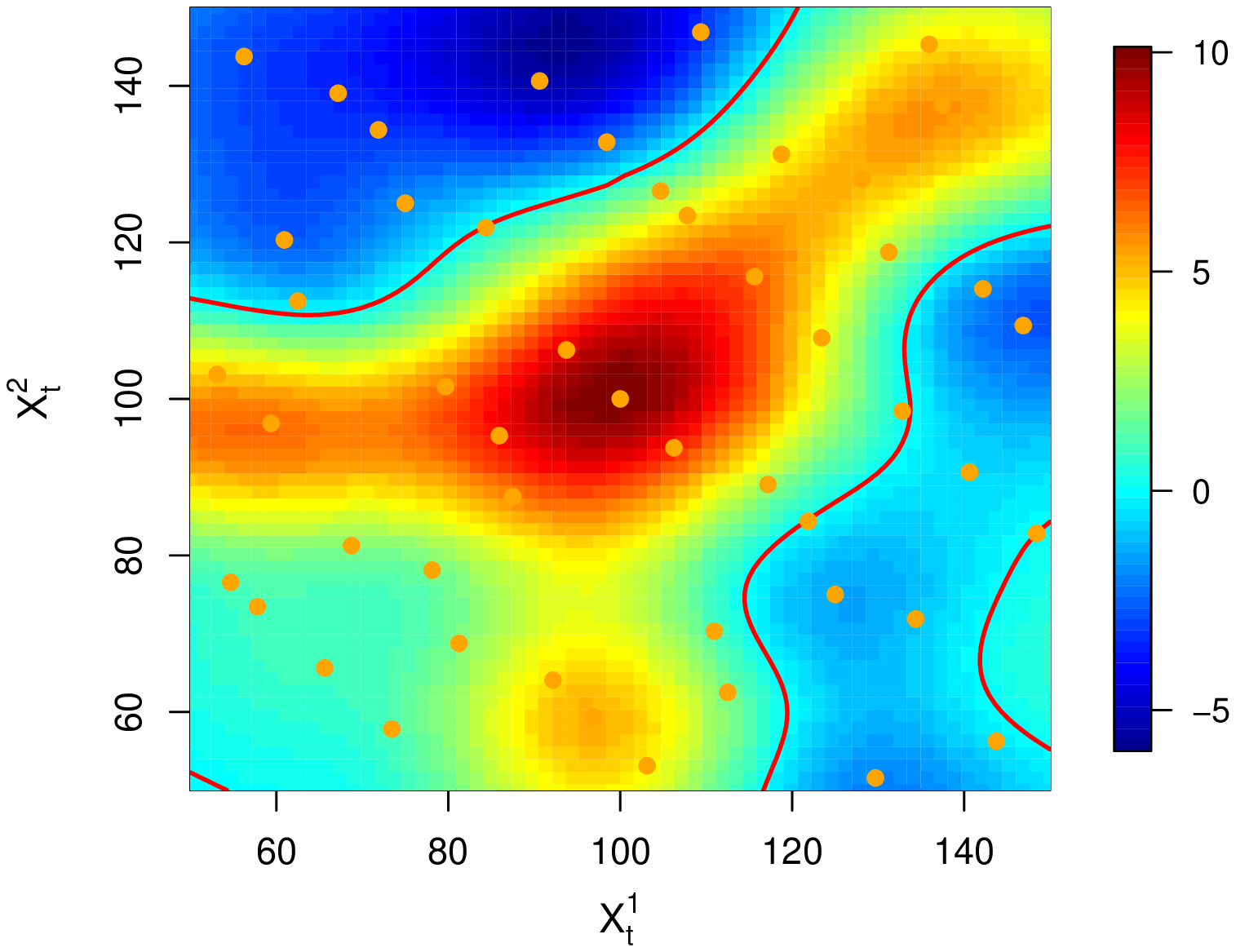} & \includegraphics[width=0.47\textwidth,height=2.5in,trim=0.1in 0.1in 0.1in 0.1in]{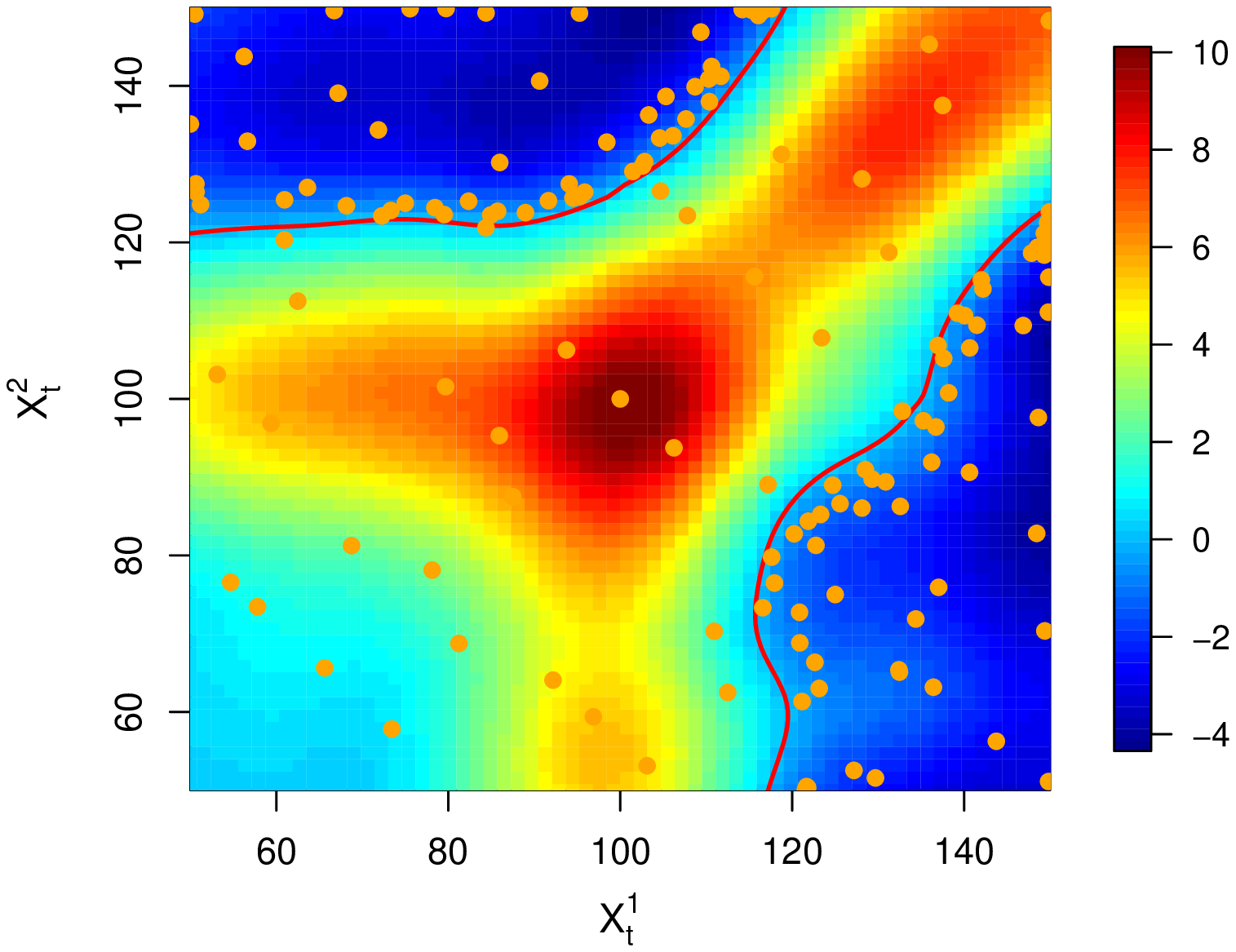} \\
 $k=50$ & $k=150$ \end{tabular}
  \caption{\red{Sequential learning of the exercise boundary $\partial \mk{S}_t$ using a ZC-SUR expected improvement criterion \eqref{eq:ei-sur} for generating $\ZZ$. We show two intermediate designs $\ZZ^{(k)}$ along with
  corresponding boundaries $\partial \mk{S}^{(k)}_t$. The underlying model is the 2D Max Call from Section \ref{sec:max-call}.}
\label{fig:sur-learning} }
\end{figure}

\red{Figure \ref{fig:sur-learning} illustrates the ZC-SUR algorithm in a 2D Max-Call setting of Section \ref{sec:max-call}. The left panel shows the initial Sobol macro design $\ZZ^{(N_0)}$ of $N_0 = 50$ sites. The right panel shows the augmented design $\ZZ^{(150)}$. At each site, a batch of $M=80$ replications is used to reduce intrinsic noise. We observe that the algorithm aggressively places most of the new design sites around the zero-contour of $\hat{\mk{S}_t}$, primarily lowering the surrogate variance (i.e.~maximizing accuracy) in that region. }

\begin{figure}[ht]
  \centering
  \includegraphics[width=0.6\textwidth,height=2in,trim=0.15in 0.15in 0.15in 0.15in]{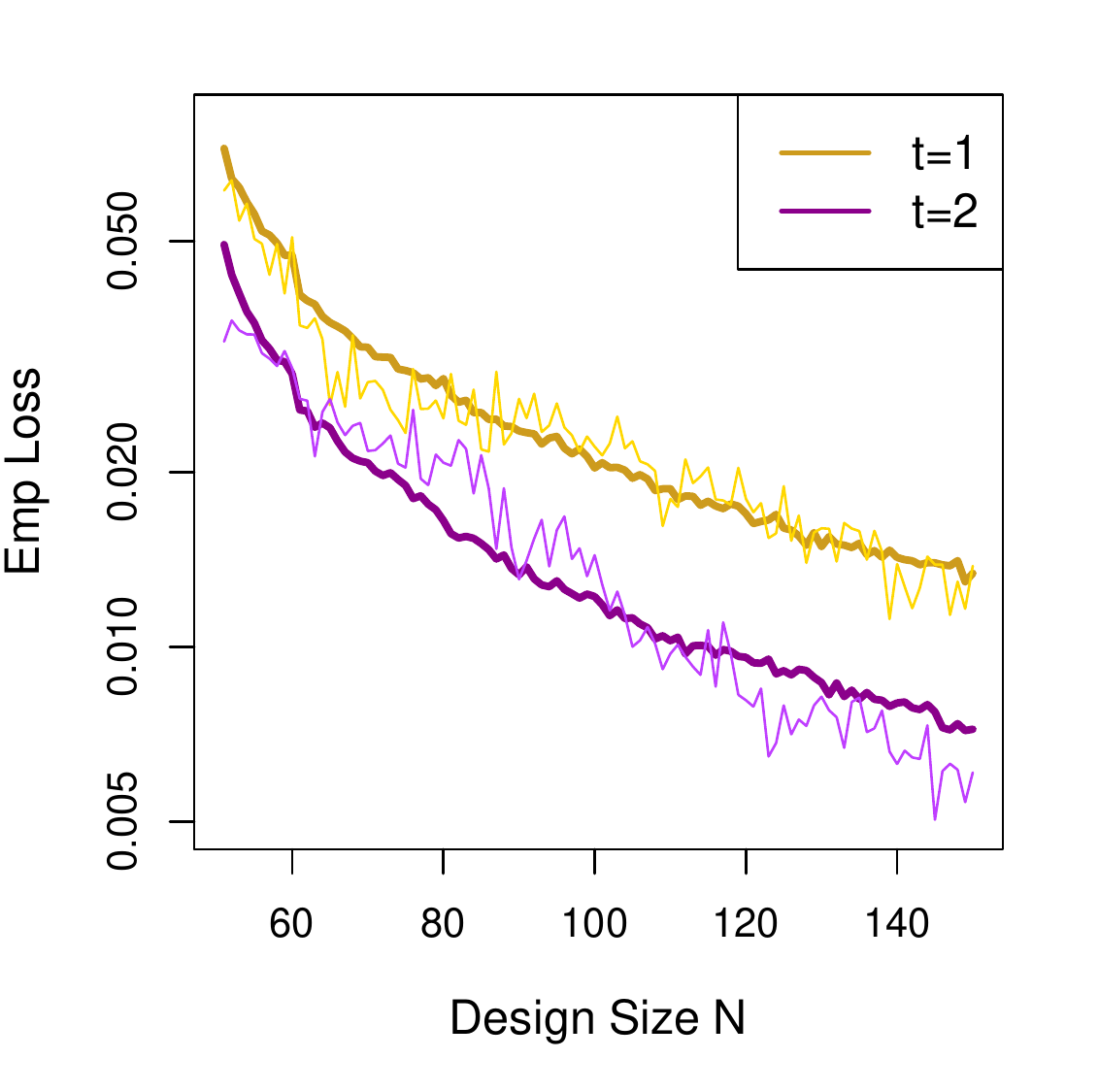}
  \caption{Evolution of $\widehat{L}^{(k)}_t$ for the 2D Max-Call problem of Section \ref{sec:max-call}. We plot the empirical integrated loss $\widehat{L}^{(k)}_t$ at $t=1$ and $t=2$, as well as the average $Ave(\widehat{L}^{(k)}_t)$ across 30 runs of Algorithms \ref{algo:rmc}-\ref{algorithm} to illustrate Monte Carlo variance and the average convergence rate of integrated loss. The sequential design utilized the ZC-SUR acquisition function \eqref{eq:ei-sur}.  \label{fig:empLoss} }
\end{figure}

The empirical integrated local loss $\widehat{L}_{RMC}$ from \eqref{eq:hat-L} provides an indicator of algorithm performance by quantifying the accuracy of $\hat{C}$. This is particularly relevant for the sequential design approach, where one is able to track $\widehat{L}_{RMC}( \hat{C}^{(k)}; \ZZ^{(k)})$ as the designs $\ZZ^{(k)}$ are augmented. Figure \ref{fig:empLoss} shows the evolution of $\widehat{L}^{(k)}_t := \widehat{L}_{RMC}(\hat{C}^{(k)}(t,\cdot); \ZZ^{(k)}_t)$ as function of the macro-design size $k$ for two different time steps $t$ and the same 2D Max-Call contract. 
We observe that $\widehat{L}^{(k)}_t$ is generally decreasing in $k$; the SUR criterion intuitively makes the latter a random walk with negative drift given by $-E_k(x^{k+1})$. Also, we observe that approximation errors are larger for later time steps, i.e.~$t \mapsto \widehat{L}^{(k)}_t$ is increasing. This is because $\ZZ_t$ is based on the underlying distribution of $X_t$; for larger $t$ the volume of the effective support of the latter grows, lowering the density of the design sites. This increases surrogate variance $v^2(x)$ and in turn raises the local loss $\ell_{RMC}(x; \ZZ^{(k)}_t)$ in \eqref{eq:emp-loss}.

While the values of $\widehat{L}^{(k)}_t$ are not sufficient on their own to yield a confidence interval for the option price (due to the complex error propagation), self-assessment allows the possibility of adaptive termination of Algorithm \ref{algo:rmc}. For example, one could implement Algorithm \ref{algo:rmc} until $\widehat{L}^{(k)}_t \le Tol$ for a pre-specified tolerance level. From above discussion, this would lead to larger experimental designs for $t$ close to $T$ which in particular would counteract error propagation.

\section{Numerical Experiments}\label{sec:examples}

We proceed to implement our RMC methods on three different types of Bermudan options in dimensions $d=2,3,5$. Our examples are based on previously published parameter sets, allowing direct comparison of relative accuracy. \red{To benchmark our approach, we compare against two implementations of the conventional LSMC, namely global polynomial bases, and local piecewise-linear bases from the method of Bouchard and Warin \cite{BouchardWarin10}. The latter consists of sub-dividing $\XX$ into $r^d$ equi-probable rectangular sub-domains and separately regressing against $\{x^1, \ldots, x^d\}$ in each cell. This strategy does not require user-specified basis function selection for \eqref{eq:ols}, although it is sensitive to the chosen number of sub-domains}. 
In each case study, to enable ``apples-to-apples'' comparison of different RMC methods, all approximations of $\mathfrak{S}_{0:T}$ are evaluated on a fixed out-of-sample set of $N_{out}=100,000$ paths of $(X_t)$.

\subsection{Benchmarked Examples}\label{sec:2d}

While our first visualizations were in 1-D, to offer more realistic setups we use as benchmarks two-dimensional and 3-dimensional models. In both cases we work with independent and identically distributed geometric Brownian motion factors $X^1,\ldots, X^d$.  As a first example we consider a two-dimensional basket Put where each asset $X^1,X^2$ follows the GBM dynamics \eqref{eq:gbm} under same parameters as in Section \ref{sec:1d}, namely $r=0.06, \delta=0,\sigma = 0.2$, $T=1,\Delta t= 0.04$. The payoff is $h(t,x) = e^{-rt} (K - \frac{X^1+X^2}{2})_+$ with $K=40$ and $X_0 = (40,40)$, leading to option value of $V(0,\vec{X}_0)= 1.461$. For our second example, we consider a three-dimensional rainbow or max-Call payoff $h(t,\vec{x}) = e^{-r t} (\max(x^1, x^2, x^3) - K)_+$. The parameters are from Andersen and Broadie~\cite{Broadie}: $r=0.05, \delta =0.1, \sigma=0.2, \vec{X}_0 = (90,90,90)$, $K=100$, $T=3$, $\Delta t=1/3$, with option value of $V(0,\vec{X}_0) = 11.25$.

\red{As a first step, Tables \ref{tbl:1d}-\ref{tbl:3d} show the impact of different experimental designs. We concentrate on the space-filling designs, evaluating the randomized LHS design, as well as two low-discrepancy sequences (Halton and Sobol), varying the batch sizes $M$. For completeness, we also compare them against the Probabilistic density-based design, and a Sequential (namely ZC-SUR) DoE strategies.  For the basket Put (see Table \ref{tbl:1d}), the designs operate with the triangular input space $\widetilde{\XX} = [25,55]^2 \cap \{ (x_1 + x_2)/2 \le K \}$ which covers the relevant in-the-money region of the contract and we use $M \in \{4,20,100\}$ (the QMC sequences were generated in a cube and then restricted to the triangular subset). For the max-Call (see Table \ref{tbl:3d}), the designs are on the rectangular region $\widetilde{\XX} =[50,150]^3$ with $M \in \{25, 80\}$. In both examples we purposely choose a very limited simulation budget ($N=3,000$ for the 2-D Put and $N=16,000$ for the 3-D Call) to accentuate the difference between the methods.}

Not surprisingly, fully capturing the shape of the exercise boundary with just a handful of design sites (even
in the idealized setting where  one obtains a noise-free sample of $C(t,x)$ at an optimal collection $x^{1:N'}$) is impossible. Nevertheless, compared to having tens of thousands of design sites in LSMC, being able to get away with a couple hundred macro-sites $N' < 200$ is almost magical. In this example we also see that the probabilistic design performs very well, partly because the Put is ATM originally. \red{At the same time the performance of an LHS design is a little bit off, and the two low-discrepancy sequences perform essentially the same.} We stress that differences of less than 0.1 cents in $\hat{V}(0,X_0)$ are essentially negligible, so that the main conclusion from Table \ref{tbl:1d} is that the precise structure of a space-filling DoE design seems to play only a minor role.

Relative to conventional LSMC, the more sophisticated kriging meta-model introduces
significant regression overhead. In standard RMC, about 95\% of simulation time is spent on path-generation and the rest on regression; in our approach this allocation is turned on its head with $> 90\%$ of time devoted to regression. This becomes even more extreme in sequential methods, where the time cost of simulations becomes essentially negligible. The regression cost arises due to the $O(N'^2)$ complexity of making predictions  \eqref{eq:krig-mean} with the kriging model; as the size $N'=N/M$ of the macro design increases, this becomes the dominant expense of the whole algorithm.  As a result, the running time of our method is very sensitive to $M$. For comparison, at $M=100$, a single run for the 2-D Put problem took 8 seconds in our computing environment, $M=20$ took 20 seconds, and $M=4$ took 190 seconds. The situation is even more acute for the sequential DoE that is an order of magnitude slower; for that reason we have not run $M=4$ (which takes many minutes) in that case. The above facts indicate that batching is essential to adequate performance of the kriging RMC. A full investigation into the optimal amount of batching (the choice of $M$ in \eqref{eq:batch-sd}) is beyond the scope of this paper. However, based on our experience, we suggest that $M\simeq 50$ would be appropriate in a typical financial context, offering a good trade-off between computational speed-up  from replication and maintaining an adequate number of distinct design sites.

\begin{table}[ht]
$$\begin{array}{lrrr}\hline
\text{Design/Batch Size} &     M=4         &   M = 20      &    M= 100         \\ \hline\hline
\text{Probabilistic} & 1.458 \ (0.002)  & 1.448 \ (0.003) & 1.443 \ (0.006)     \\
\text{LHS} & 1.453 \ (0.002)  & 1.446 \ (0.004) & 1.416 \ (0.033)     \\
\text{Sobol QMC} & 1.454 \ (0.002)& 1.448\  (0.002) & 1.454\  (0.002)     \\
\text{\red{Halton QMC}}  &  1.456\ (0.003)  & 1.447\ (0.003)  &  1.452\ (0.003) \\
\text{Sequential ZC-SUR} &  N/A              & 1.438\  (0.003) & 1.450\  (0.003)     \\ \hline
\end{array}$$
\caption{Performance of different DoE approaches to RMC in the 2-D Bermudan Put setting of Section \ref{sec:2d}. All methods utilize $|\ZZ_t | = 3000$. Results are averages (standard deviations) of 100 runs of each method,  evaluating $\hat{V}(0,X_0)$ on a fixed out-of-sample database of $N_{out} = 10^5$ scenarios. For comparison, LSMC-BW11 algorithm yielded estimates of $\hat{V}^{BW11}(0,X_0)=1.431$ with $N=10,000$ and $\hat{V}^{BW11}(0,X_0)=1.452$ with $N=50,000$.\label{tbl:1d}}
\end{table}

\begin{table}
  $$\begin{array}{lrr}\hline
\text{Design/Batch Size}  &   M = 25      &    M= 80         \\ \hline\hline
\text{Probabilistic} & 11.115 \ (0.015)  & 11.128 \ (0.015)     \\
\text{LHS} & 11.071 \ (0.025)  & 11.110 \ (0.029)      \\
\text{Sobol QMC} & 11.131 \ (0.020)& 11.177\  (0.015)      \\
\text{\red{Halton QMC}}  & 11.120 \ (0.017)  &  11.183 \ (0.017)   \\
\text{Sequential ZC-SUR} &  11.160 \ (0.025)    & 11.147\  (0.014)     \\ \hline
\end{array}$$
\caption{Performance of different DoE approaches to RMC in the 3-D Bermudan Max-Call setting of Section \ref{sec:2d}. All methods utilize $|\ZZ_t | = 16,000$. Results are averages (standard deviations) of 100 runs of each method, with $\hat{V}(0,X_0)$  estimated from a fixed out-of-sample database of $N_{out} = 10^5$ scenarios. For comparison, LSMC-BW11 algorithm yielded estimates of $\hat{V}^{BW11}(0,X_0)=11.07$ with $N=10^5$ and $\hat{V}^{BW11}(0,X_0)=11.12$ with $N=3 \cdot 10^5$.\label{tbl:3d}}
\end{table}

Table \ref{tbl:kernel} illustrates the effect of choosing different kriging kernels $\cK$; we consider the Matern-5/2 \eqref{eq:matern}, Gaussian/squared-exponential \eqref{eq:sqexp} and Matern-3/2 families \cite{kmPackage-R}. In each case kernel hyper-parameters are fitted in a black-box manner by the \texttt{DiceKriging} software and underlying design was space-filling using the Sobol sequence. As expected, optimizing the kernel gives more accurate results, improving the previous, fixed-kernel answers in  Tables \ref{tbl:1d}-\ref{tbl:3d} above; it also adds about 20\% additional time. The other conclusion is that the choice of the kernel family has a negligible effect on performance. Hence, for the rest of the experiments we work with the default choice of a squared-exponential kernel.

\begin{table}
\red{  $$\begin{array}{lrr}\hline
\text{Kernel}  &  \text{2-D Basket Put}     &   \text{3-D Max Call}         \\ \hline\hline
\text{Matern-5/2} & 1.452 \ (0.00243)  &  11.191 \  (0.0139)     \\
\text{Matern-3/2} & 1.450 \ (0.00252)  & 11.191\ (0.0136) \\ 
\text{Gaussian} & 1.453 \ (0.00273)&  11.172\ (0.0160) 
\end{array}$$
\caption{Performance of different kriging models for the 2D Put and the 3D Call. The table reports $\hat{V}(0,X_0)$ and its Monte Carlo standard deviation (in parentheses) across algorithm runs. Design sizes were $|\ZZ|=3,000$ in 2-D and $| \ZZ |=16,000$ in 3-D and used the Sobol QMC design. Results are based on averaging 100 runs of each method, and evaluating $\hat{V}(0,X_0)$ on a fixed out-of-sample database of $N_{out} = 100,000$ scenarios. \label{tbl:kernel}} }
\end{table}

\red{From the above experiments, the following guidance can be given for implementing kriging RMC: pick $M$ in the range 50--100 which is sufficient to estimate $\sigma^2(x)$ reasonably well, and reduce macro design size $N'$ significantly; use whatever kernel family and hyper-parameter optimization the software recommends. As a goodness-of-fit check, examine posterior variance $v^2(x)$ along the estimated stopping boundary $\partial \hat{\mk{S}}_t$ ---it should not be too large.}

\subsection{Scalability in State Dimension}\label{sec:max-call}
\red{The max-Call setting is convenient for investigating scalability of our approaches in the dimension $d$ of the problem. In contrast to basket Put, the stopping region of a max-Call consists of several disconnected pieces. Namely, it is optimal to stop if exactly one asset is in-the-money $X^1 > K$ while all other assets are below the strike $K$. This generates
$d$ disconnected stopping sub-regions in dimension $d$. One consequence is that selecting a good domain  $\widetilde{\XX}$ for a space-filling design, such as one of the low-discrepancy sequences, is difficult since the continuation region is non-convex and the stopping region is disconnected.  We use $\widetilde{\XX} = [50, 150]^d$ which encompasses the relevant parts of both stopping and continuation  regions; hence, in this situation we build our metamodel for both in-the-money and out-of-the-money paths. The latter $\widetilde{\XX}$ is rather large, hurting the accuracy of space-filling designs.}

Following \cite{Broadie} we benchmark for $d=2,3,5$.
Table \ref{table:maxCall} shows the results from two different implementations of the LSMC algorithm (namely a global polynomial (``Poly'') basis, as well as the localized
Bouchard-Warin (``BW11'') \cite{{BouchardWarin10}} scheme, and comparing them to two kriging metamodels, one using a space-filling design utilizing the low-discrepancy Sobol sequence (``Krig+Sobol'') and the other a sequential ZC-SUR design (``Krig+SUR''). To show the advantage of our approach, we use an order-of-magnitude smaller design for Krig+Sobol, and an even smaller one for the adaptive Krig+SUR method. For large $d$, the standard LSMC approach requires hundreds of thousands of $X$-trajectories which may impose memory constraints, so these savings are impressive.

 In terms of overall simulation budget measured by the number of 1-step simulations of $X$ (see fourth column of Table \ref{table:maxCall}), the use of better experimental designs and kriging meta-models enables a factor of 2-5 savings, which is smaller than the reduction in $N$ because we do not exploit the ``global grid'' trick of the conventional approach. In the last five-dim.~example we also see the limitation of the kriging model due to excessive regression overhead as the macro-design size $N'=N/M$ rises. As a result, despite using less than 20\% of the simulation budget, the overall running time, is much longer than for LSMC. With the present implementation using \textbf{DiceKriging}, this overhead crowds out simulation savings around $N' > 500$.
 Table \ref{table:maxCall} suggests that while attractive conceptually, the gains from a sequential design strategy are marginal in terms of memory, and are expensive in terms of additional overhead. Rephrasing, one can squeeze most of the savings via a space-filling design. This observation offers a pragmatic guidance for the DoE aspect, indicating that one must balance simulation-budget savings against regression/DoE overhead. Of course, this trade-off keeps changing as more efficient large-scale kriging/RSM software is developed/used (e.g.~localized kriging, see \cite{gramacy:apley:2013}).

\begin{table}[ht]
  $$ \begin{array}{p{1.8in}rrrr} \\ \hline
  \text{ Method } & \hat{V}(0,X_0) & (\text{StDev.}) & \# \text{Sims} & \text{Time (secs)} \\ \hline
  \multicolumn{5}{c}{\textbf{2D Max call}} \\ \hline
  \text{\red{LSMC Poly}} \hfill  N=50,000 & 7.93 & (0.018) & 3.60 \cdot 10^5 & 3.9 \\
  \text{LSMC BW11} \hfill  N=50,000 & 7.89 & (0.023) & 3.60 \cdot 10^5 & 4.0 \\ 
  \text{Krig + Sobol} \hfill N=10,000 & 8.12 & (0.019) & 2.54 \cdot 10^5  &5.3 \\
  \text{Krig + SUR} \hfill N=6,000 & 8.10 & (0.024) & 1.44 \cdot 10^5 & 10.8 \\ \hline \hline
   \multicolumn{5}{c}{\textbf{3D Max Call}} \\ \hline
    \text{\red{LSMC Poly}} \hfill N=300,000 & 10.95 & (0.01) & 2.70 \cdot 10^6 & 15 \\
    \text{LSMC BW11} \hfill N=300,000 & 11.12 & (0.01) & 2.70 \cdot 10^6 &  22 \\  
  \text{Krig + Sobol} \hfill N=20,000  & 11.18 & (0.02) & 0.48 \cdot 10^6 & 33 \\ 
  \text{Krig + SUR} \hfill N=16,000 & 11.15 & (0.02) & 0.51 \cdot 10^6 & 161 \\ 
  \hline \hline
    \multicolumn{5}{c}{\textbf{5D Max Call}} \\ \hline
    \text{\red{LSMC Poly}} \hfill N=800,000 & 15.81 & (0.01) & 7.20 \cdot 10^6 &  42 \\
    \text{LSMC BW11} \hfill N=640,000 & 16.32 & (0.02) & 5.76 \cdot 10^6 &  87 \\  
  \text{Krig + Sobol} \hfill N=32,000  & 16.31 & (0.02) & 0.86 \cdot 10^6 & 205 \\
  \text{Krig + SUR} \hfill N=25,000 & 16.30 & (0.02) & 0.66 \cdot 10^6 & 666 \\
  \hline \hline
\end{array}$$
 \caption{Comparison of RMC methods for different max-Call models.  Results are averages across 100 runs of each algorithm, with third column reporting the corresponding standard deviations of $\hat{V}(0,X_0)$. Time is based on running the \texttt{R} code on a 1.9 MHz laptop with 8Gb of RAM. \red{The LSMC Polynomial method used a tensor product basis of degree 3 (in 2-D and 3-D) and degree 2 in 5-D.  The LSMC BW11 method used $10^2$ partitions for $d=2$, $5^3$ partitions for $d=3$ and $4^5$ partitions for $d=5$. Kriging models used the squared-exponential kernel.} \label{table:maxCall}}
\end{table}

\subsection{Stochastic Volatility Examples}\label{sec:heston2d}

Lastly, we discuss stochastic volatility (SV) models.  Because there are no (cheap) exact methods to simulate the respective asset prices, discretization methods such as the Euler scheme are employed. This makes this class a good case study on settings where simulation costs are more significant.
Our case study uses the mean-reverting SV model with a two-dimensional state $X=(X^1,X^2)$: 
\begin{align} \label{eq:SV} \left\{ \begin{aligned}
  dX^1_t &= r X^1_t \, dt + e^{X^2_t} X^1_t \, dW^1_t, \\
  dX^2_t &=  a( m_1 - X^2_t) \,dt + \nu  \, dW^2_t. \end{aligned} \right.
\end{align}
where $X^1$ is the asset price and $X^2$ the log-volatility factor. Volatility follows a mean-reverting stationary Ornstein-Uhlenbeck process with mean reversion rate $a$ and base-level $m_1$, with a leverage effect expressed via the correlation $d \langle W^1, W^2 \rangle_t = \rho dt$. Simulations of $(X^1,X^2)$ are done based on an Euler scheme with a time-step $\delta t$.
We  consider the asset Put $h(t,x^1, x^2) =  e^{-r t}(K - x^1)_+$; exercise opportunities are spaced out by $\Delta t \gg \delta t$. Related experiments have been carried out in \cite{Rambharat11}, and recently in \cite{AgarwalSircar15}.

Table \ref{tbl:2d} lists three different parameter sets.
Once again we consider the kriging meta-models (here implemented with LHS for a change) against LSMC implementation  based on the BW11 \cite{BouchardWarin10} algorithm.
Table \ref{tbl:2d} confirms that the combination of kriging and thoughtful simulation design allows to reduce $N$ by an order of magnitude. Compare for the first parameter set a kriging model with LHS design $\ZZ$ that uses $N=10,000$ (and estimates option value at 16.06) relative to the LSMC model with $N=128,000$ (and estimates option value of 16.05). For the set of examples from \cite{AgarwalSircar15} we observe that the space-filling design does not perform as well, which most likely is due to having the inefficient rectangular LHS domain $\widetilde{\XX} = [30,50] \times [-4,1]$. Still, the sequential design method wins handily.

\begin{table}[ht]
  $$ \begin{array}{p{1.8in}rrrr} \\ \hline
  \text{ Method } & \hat{V}(0,X_0) & (\text{StDev.}) & \# \text{ of Sims} & \text{Time (secs)} \\ \hline
  \multicolumn{5}{c}{\textbf{Rambharat \& Brockwell \cite{Rambharat11} Case SV5}} \\
  \multicolumn{5}{c}{r=0.0225, a=0.015, m=2.95, \nu=3/\sqrt{2}, \rho=-0.03} \\
  \multicolumn{5}{c}{K=100, X^1_0 = 90, X^2_0 = \log 0.35, T=50/252, \Delta t= 1/252, \delta t= 1/2520} \\
   \hline
  \text{LSMC BW11} \hfill N=48,000 & 15.89 & (0.08) & 2.50 \cdot 10^6 & 24 \\ 
  \text{LSMC BW11} \hfill N=128,000 & 16.03 & (0.05) & 6.25 \cdot 10^6 & 52 \\ 
  \text{Krig + LHS} \hfill N=4000 & 15.58 & (0.22) & 0.96 \cdot 10^6 & 25 \\
  \text{Krig + LHS} \hfill N=10,000 & 16.06 & (0.05) & 4.25\cdot 10^6 & 155 \\
  \text{Krig + SUR} \hfill N=4000 & 16.24 & (0.04) & 2.32 \cdot 10^6 & 283 \\
  \text{Krig + SUR} \hfill N=10,000 & 16.30 & (0.05) & 5.69 \cdot 10^6 & 565 \\ \hline \hline
   \multicolumn{5}{c}{\textbf{Agarwal, Juneja \& Sircar ITM  \cite{AgarwalSircar15} $X^1_0 = 90$}} \\
     \multicolumn{5}{c}{r=0.1, a=1, m=-2, \nu=\sqrt{2}, \rho=-0.3} \\
  \multicolumn{5}{c}{K=100, X^1_0 = 90, X^2_0 = -1, T=1, \Delta t= 0.05, \delta t= 0.001} \\ \hline
    \text{LSMC BW11} \hfill N=50,000 & 14.82 & (0.03) & 1.00 \cdot 10^6 &  36 \\  
  \text{LSMC BW11} \hfill N=125,000 & 14.95 & (0.02) & 2.50 \cdot 10^6 & 82 \\ 
  \text{Krig + LHS} \hfill N=5000 & 14.73   & (0.05) & 0.26 \cdot 10^6 & 13 \\
  \text{Krig + LHS} \hfill N=20,000 & 14.81 & (0.02) & 1.05\cdot 10^6 & 85 \\
    \text{Krig + SUR} \hfill N=4000 & 14.91 & (0.03) & 0.25 \cdot 10^6  & 50 \\ 
  \text{Krig + SUR} \hfill N=10,000 & {14.96} & (0.02) & 0.58 \cdot 10^6 & 121 \\\hline
    \multicolumn{5}{c}{\textbf{Agarwal, Juneja \& Sircar OTM \cite{AgarwalSircar15} $X^1_0= 110$}} \\ \hline
    \text{LSMC BW11} \hfill N=50,000 & 8.25 & (0.02) & 1.25 \cdot 10^6 &  26.7 \\  
  \text{LSMC BW11} \hfill N=125,000 & 8.30 & (0.02) & 3.12 \cdot 10^6 & 67.3 \\ 
  \text{Krig + LHS} \hfill N=5000 & 8.27 & (0.02) & N/A & N/A \\
  \text{Krig + LHS} \hfill N=20,000 & 8.30 & (0.01) & N/A & N/A \\ \hline \hline
  \end{array}$$
  \caption{Comparison of RMC methods for Bermudan Puts under stochastic volatility models \eqref{eq:SV}. Results are averages across 100 runs of each algorithm, with third column reporting the corresponding standard deviations of $\hat{V}(0,X_0)$. \label{tbl:2d}}
\end{table}

Table \ref{tbl:2d} also showcases another advantage of building a full-scale metamodel for the continuation values. Because we obtain the full map $x \mapsto \hat{C}(t,\cdot)$, our method can immediately generate an approximate exercise strategy for a range of initial conditions. In contrast, in conventional LSMC all trajectories start from a fixed $X_0$. In Table \ref{tbl:2d} we illustrate this capability by re-pricing the Put from \cite{AgarwalSircar15} after changing to an out-of-the-money initial condition $X^1_0 = 110$ (keeping $X^2_0 = -1$). For the kriging approach we use the exact same metamodels that were built for the ITM case ($X^1_0 = 90$), meaning the policy sets $\widehat{\mathfrak{S}}_{0:T}$ remain the same. Thus, all that was needed to obtain the reported Put values $\hat{V}(0,X_0)$ was a different set of test trajectories to generate fresh out-of-sample payoffs. \red{We note that the obtained results (compare $\hat{V}^{Krig}(0,X_0)=8.27$ with budget of $N=5000$ to $\hat{V}^{LSMC}(0,X_0)=8.25$ with $N=50,000$) still beat the re-estimated LSMC solution by several cents.}

\section{Conclusion}\label{sec:conclude}

We investigated the use of kriging meta-models for policy-approximation based on \eqref{eq:hatS-t} for optimal stopping problems. \red{As shown, kriging allows a flexible modeling of the respective continuation value,
internally learning the spatial structure of the model. Kriging is also well-suited for a variety of DoE approaches, including batched designs that alleviate non-Gaussianity in noise distributions, and adaptive designs that refine local accuracy of the meta-model in the regions of interest (and demand a localized, non-parametric fit). These features translate into savings of 50-80\% of simulation budget. While the time-savings are often negative, I believe that kriging is a viable, and attractive regression approach for RMC. First, in many commercial-grade applications, memory constraints are as important as speed constraints. Speed is generally much more scalable/parallelizable than memory. Second, as implemented, the main speed bottleneck comes from $O( (N')^3)$ cost of building a kriging model; alternative formulations get around this scalability issue, and more speed-ups can be expected down the pike.  Third, kriging brings a suite of diagnostic tools, such as a transparent credible interval for the fitted $\hat{C}$ that can be used to self-assess accuracy. Fourth, through its analytic structure kriging is well suited for automated DoE strategies.}

Yet another advantage of kriging, is a consistent probabilistic framework for joint modeling of $\hat{C}$ and its derivatives. The latter is of course crucial for \emph{hedging}: in particular Delta-hedging is related to the derivative of the value function with respect to $x$. In the continuation region we have $\partial_x V(t,\cdot) = \partial_x C(t,\cdot)$, so that a natural estimator for Delta is the derivative of the meta-model surrogate $\hat{\Delta}(t,x) = \partial_x \hat{C}(t,x)$, see e.g.~\cite{WangCaflisch09,JainOosterlee13}. For kriging metamodels, the posterior distribution of the response surface derivative $\partial_x \hat{C}(t,\cdot)$ is again Gaussian and thus available analytically, with formulas similar to \eqref{eq:krig-mean}-\eqref{eq:krig-cov}. We refer to \cite{AnkenmanNelson13} for details, including an example of applying kriging for computing the Delta of European options. Application of this approach to American-style contracts will be explored separately.

\red{The presented meta-modeling perspective modularizes DoE and regression, so that one can mix and match each element. Consequently, one can swap kriging with another strategy, or alternatively use a conventional LSMC with a space-filling design. In the opinion of the author, this is an important insight, highlighting the range of choices that can/need to be made within the broader RMC framework. In a similar vein, by adding a separate pre-averaging step that resembles a nested simulation or Monte Carlo forest scheme (see \cite{BenderKolodko08}), our proposal for batched designs highlights the distinction between smoothing (removing simulation noise) and interpolation (predicting at new input sites), which are typically lumped together in regression models. To this end, kriging meta-models admit a natural transition between modeling of deterministic (noise-free) and stochastic simulators and are well-suited for batched designs. }

\red{Kriging can also be used to build a classification-like framework that swaps out cross-sectional regression to fit $\hat{C}$ with a direct approach to estimating $\hat{\mk{S}}$.  One solution is to convert the continuous-valued path payoffs $H_t(x^n_\cdot)$ into labels: $Z_t(x) := 1(H_t(x^n_\cdot) < h(t,x))$ and build a statistical classification model $\hat{p}_t$ for $p_t(x) := \PP( Z_t(x) = 1 | x)$; for example a kriging probit model. Then $\hat{\mk{S}}_t := \{ x : \hat{p}_t(x) > 0.5\}$. Modeling $Z$ rather than $Y$ might alleviate much of the ill-behavior in the observation noise $\eps$. These ideas are left for future work.
}

By choice, in our experiments we used a generic DoE strategy with minimal fine-tuning. In fact, the DoE perspective suggests a number of further enhancements for implementing RMC. Recall that the original optimal stopping formulation is reduced to iteratively building $T$ meta-models across the time steps. These meta-models are of course highly correlated, offering opportunities for ``warm starts'' in the corresponding surrogates. The warm-start can be used both for constructing adaptive designs $\ZZ_t$ (e.g.~a sort of importance sampling to preferentially target the exercise boundary of $\partial\mk{S}_{t+1}$) and for constructing the meta-model (e.g.~to better train the kriging hyper-parameters $(s^2,\vec{\theta})$). Conversely, one can apply different approaches to the different time-steps, such as building experimental designs of varying size $N_t$, or shrinking the surrogate domain $\widetilde{\XX}_t$ as $t \to 0$. These ideas will be explored in a separate forthcoming article.

As mentioned, ideally the experimental design $\ZZ$ is to be concentrated along the stopping boundary. While this boundary is a priori unknown, some partial information is frequently available (e.g.~for Bermudan Puts one has rules of thumb that the stopping boundary is moderately ITM). This could be combined with an importance sampling strategy to implement the path-generation approach of standard LSM while generating a more efficient design. See \cite{DelMoralHu12,JunejaKalra09,GobetTurkedjiev15} for various aspects of importance sampling for optimal stopping. \red{Another variant would be a ``double importance sampling'' procedure of first generating a non-adaptive (say density-based) design, and then adding (non-sequentially)  design points/paths in the vicinity of estimated $\partial \hat{\mk{S}}_t$.}

\medskip
\textbf{Acknowledgment}: I am grateful to the anonymous referee and the editor for many useful suggestions which have improved the final version of this work. 

\bibliographystyle{siam}
\bibliography{../../../masterBib}

\end{document}